\documentclass[a4paper,12pt]{article}

\usepackage{amssymb,amsmath,epsf,graphicx}

\newcommand{\im}{\mathrm{Im\,}}
\newcommand{\tri}[4]{#1_{#2 #3}^{\phantom{#2} #4}}
\newcommand{\uhp}{\mathbb{H}^+}
\newcommand{\disc}{\mathbb{D}}
\newcommand{\cor}[1]{\left\langle #1 \right\rangle}
\newcommand{\coruhp}[1]{\left\langle #1 \right\rangle_{\mathbb{H}^+}}
\newcommand{\cordisc}[1]{\left\langle #1 \right\rangle_{\mathbb{D}}}
\newcommand{\id}{{1\!\!1}}
\newcommand{\ishiket}[1]{|#1\rangle\!\rangle}
\newcommand{\braneket}[1]{\|#1\rangle\!\rangle}
\allowdisplaybreaks[2]

\begin{document}

\vspace*{-1.5cm}
\thispagestyle{empty}
\begin{flushright}
AEI-2009-063
\end{flushright}
\vspace*{2.5cm}

\begin{center}
{\Large 
{\bf Bulk flows in Virasoro minimal models with boundaries}}
\vspace{2.5cm}

{\large Stefan Fredenhagen\footnote[1]{{\tt E-mail: stefan@aei.mpg.de}},
Matthias R.\ Gaberdiel\footnote[2]{{\tt E-mail: gaberdiel@itp.phys.ethz.ch}} 
and Cornelius Schmidt-Colinet\footnote[3]{{\tt E-mail: schmidtc@itp.phys.ethz.ch}}} 
\vspace*{0.5cm}

$^{1}$Max-Planck-Institut f{\"u}r Gravitationsphysik,
Albert-Einstein-Institut\\ 
D-14424 Golm, Germany\\
\vspace*{0.5cm}

$^{2,3}$Institut f{\"u}r Theoretische Physik, ETH Z{\"u}rich\\
CH-8093 Z{\"u}rich, Switzerland\\
\vspace*{3cm}

{\bf Abstract}
\end{center}
The behaviour of boundary conditions under relevant bulk perturbations
is studied for the Virasoro minimal models. In particular, we consider
the bulk deformation by the least relevant bulk field which
interpolates between the $m^{\rm th}$ and $(m-1)^{\rm st}$ unitary
minimal model.  In the presence of a boundary this bulk flow induces
an RG flow on the boundary, which ensures that the resulting boundary
condition is conformal in the $(m-1)^{\rm st}$ model. By combining 
perturbative RG techniques with insights from defects and results
about non-perturbative boundary flows, we determine the endpoint of the flow,
{\it i.e.}\ the boundary condition to which an arbitrary boundary
condition of the $m^{\rm th}$ theory flows to.

\newpage
\renewcommand{\theequation}{\arabic{section}.\arabic{equation}}

\section{Introduction}

Perturbations of conformal field theories by marginal or relevant operators
play an important role in various contexts, for example in string theory where
they describe string moduli or time dependent processes. Many aspects 
of perturbed conformal field theories have been studied over the years, 
starting from the seminal work of Zamolodchikov 
\cite{Zamolodchikov:1987ti,Zamolodchikov:1989zs}, and a number of 
structural results are known, in particular the $c$-theorem  for bulk
perturbations \cite{Zamolodchikov:1986gt}, as well as the $g$-theorem 
for boundary perturbations \cite{Affleck:1991tk,Friedan:2003yc}. 

Most of the work so far has been done on bulk perturbations of bulk conformal
field theories, or boundary perturbations of boundary conformal field theories. 
However, it is clear that a bulk perturbation will also affect the boundary condition
since the boundary condition of the original theory will typically not be conformal
with respect to the new bulk conformal fixed point. This combined problem has only 
recently been addressed from the point of view
of perturbed conformal field theory \cite{Fredenhagen:2006dn,FGK2,Gaberdiel:2008fn}
(see also \cite{Gaberdiel:2007us,Gaberdiel:2008rk}), although there has been
earlier work in the context of integrable models starting from \cite{Cherednik:1985vs} 
and further developed in \cite{Sklyanin:1988yz,Fring:1993mp,Ghoshal:1993tm}. 
In particular, these flows have been studied 
using a version of the thermodynamic Bethe ansatz (see for example
\cite{LeClair:1995uf,Lesage,Dorey:1997yg,Dorey:1999cj,Dorey:2004}),
the truncated conformal space approach (see for example 
\cite{Dorey:1997yg,Dorey:1999cj,Dorey:2000}) and a form factor
expansion \cite{Bajnok:2006ze,Castro-Alvaredo:2006sh}. From the point of 
view of perturbed conformal field theory, 
a bulk perturbation generically induces a boundary renormalisation
group (RG) flow that will ensure that at the endpoint of the flow both
bulk and boundary are again  
conformal~\cite{Fredenhagen:2006dn,FGK2,Gaberdiel:2008fn}. 

The resulting coupled RG equations have so far only been worked out for a few 
simple examples. In all of them, the bulk perturbation was actually exactly
marginal in the bulk. As a consequence the bulk RG equation was trivial, 
and one only had to solve the boundary RG equation with a (bulk) source
term. In this paper we shall demonstrate that these techniques also work nicely
for a genuinely coupled bulk boundary problem, where neither of the
perturbations is marginal.
\smallskip

The archetypal examples for which these kinds of problems can be studied are 
the Virasoro minimal models. Indeed, the original analysis of Zamolodchikov
\cite{Zamolodchikov:1987ti} was performed in this context: he 
established that the perturbation of the $m^{\rm th}$ minimal model by the least
relevant field, the bulk field $\phi_{(1,3)}$, induces an RG flow whose endpoint is 
the $(m-1)^{\rm st}$ minimal model. The analogous analysis for the boundary
perturbation --- the perturbation of a Cardy boundary condition \cite{Cardy:1989ir}
of the $m^{\rm th}$ minimal model by the $\psi_{(1,3)}$ boundary field --- was
done in \cite{Recknagel:2000ri}. They showed that the endpoint of this boundary flow
is in general a superposition of fundamental boundary conditions. For the combined
problem, the bulk perturbation by $\phi_{(1,3)}$ in the presence of a boundary, 
only a few numerical studies have been performed so far
\cite{Pearce:2000dv,Pearce:2003km}, and some conjectural TBA results
exist~\cite{Lesage,Ahn:1998xm}. In this paper we will fill in this gap, 
and show how this problem can be analysed analytically. 

Let us briefly sketch our argument. The bulk RG equation
is unaffected by the presence of the boundary, and thus the old fixed point analysis
of Zamolodchikov applies. The boundary RG equation, on the other hand, is of the form
\begin{equation}
\dot{\mu}=(1-h)\mu+\tfrac{1}{2}B\, \lambda+ D\, \mu^2 + E \, \lambda \mu + F \lambda^2 + 
{\cal O}(\mu^3, \lambda\mu^2, \lambda^2\mu,\lambda^3) \ .
\end{equation}
Here the first and third term are the usual boundary RG equation terms
for the boundary coupling $\mu$, while the other three terms involve
also the bulk coupling constant $\lambda$. The $B\lambda$ term is the source
term that was studied in \cite{Fredenhagen:2006dn}, whereas the
$E\lambda\mu$ term describes how the bulk deformation modifies the
conformal weight of the boundary field \cite{Gaberdiel:2008fn}. In
some sense this term only appears at higher order in perturbation
theory, and it was only recently understood how to calculate it as an
integral of a chiral four-point function \cite{Gaberdiel:2008fn}. The
$F\lambda^2$ term has not so far been studied in detail, but the
coefficient $F$ itself is parametrically small --- it is of order
$1/m$ --- and thus the $F\lambda^2$ term is subleading. In fact, our
analysis will be performed for large $m$, for which we shall find
perturbative fixed points for $\lambda$ and $\mu$ that are of order
$1/m$. The $E\lambda\mu$ term is then of the same order as the
standard $D\mu^2$ term, while the $F\lambda^2$ will be subleading and
can be ignored to leading order. 

The resulting RG equations have generically three perturbative fixed points:
the pure boundary perturbation fixed point of \cite{Recknagel:2000ri}, as well 
as two perturbative fixed points in the $(m-1)^{\rm st}$ theory ({\it i.e.}\ for
non-trivial $\lambda$). The first fixed point (I)  can be identified as in 
\cite{Recknagel:2000ri}, namely by computing the perturbed $g$-function
and identifying it with the $g$-function of the fixed point boundary condition. 
However, the analysis for the other two fixed points (II \& III) is not so 
straightforward since they live in a different bulk theory than the one 
we started with, and 
it is therefore not clear to which extent we can compare the 
$g$-functions directly.  However, it is reasonable to assume that it makes
sense to compare  
ratios of $g$-functions \cite{Dorey:1999cj}, and it is furthermore plausible that 
the boundary condition with the overall smallest $g$-function --- this is the boundary 
condition that corresponds to the identity representation --- should flow to the 
corresponding boundary condition in the $(m-1)^{\rm st}$ theory. This tells us how 
the overall scale of the $g$-functions changes, and thus allows us to make 
a definite prediction for the $g$-function of the fixed point in the  $(m-1)^{\rm st}$ 
minimal model. 
\begin{figure}[tb]\begin{center}
\includegraphics{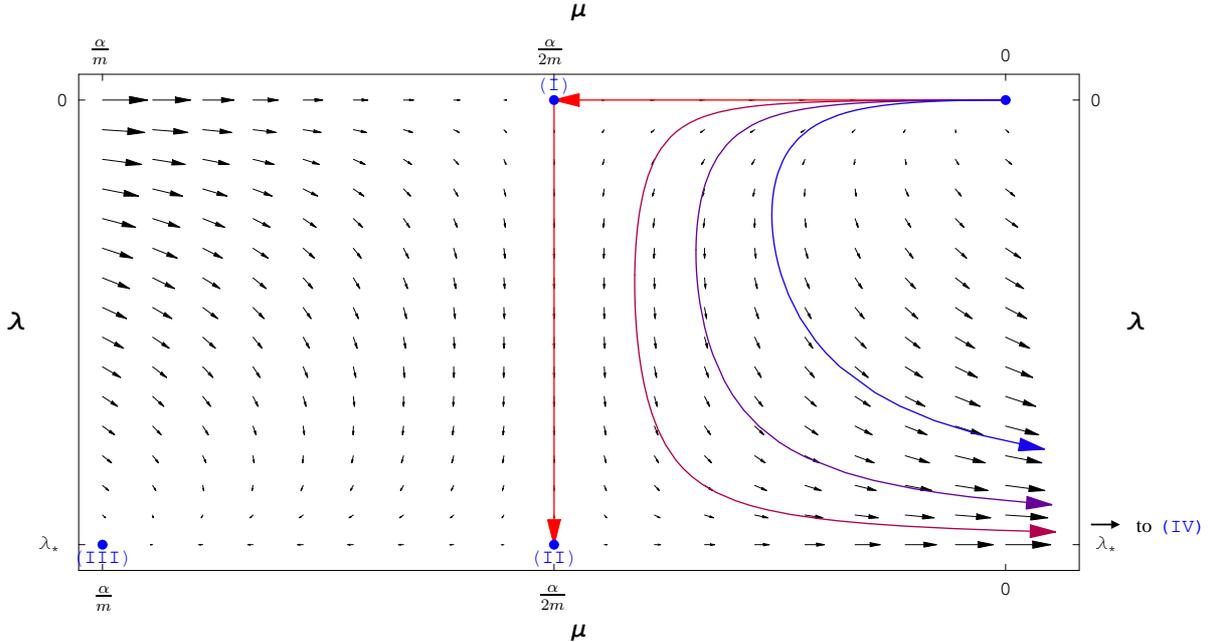}
\end{center}
\caption{\label{fig:flowgraph}The combined flow diagram for $(a_{1},a_{2})=(2,3)$ and
$m=100$ (for which $\alpha=-4$;  see section~2.2 for details). 
We have magnified  the vectors 
$(\dot{\mu},\dot{\lambda})$ 
by a factor $2.5$. The horizontal arrow indicates the pure boundary
flow to the perturbative fixed-point (I) in the $m^{\rm th}$ minimal model, 
the vertical arrow describes the flow of the boundary
condition (I) to the boundary condition (II) in the $(m-1)^{\rm st}$ minimal
model. The three other flows that are depicted are generic
bulk-boundary RG-flows that at the end tend towards 
the fixed point (IV) where $\mu =+\infty$ in
the $(m-1)^{\rm st}$ minimal model ($\lambda =\lambda_{*}$).}
\end{figure}
Progressing in this manner, we can then identify the two
perturbative fixed points in the $(m-1)^{\rm st}$ theory. As it turns out, 
one of the fixed points (III) is actually the end-point of a pure  $\psi_{(1,3)}$ 
boundary perturbation of the other (II) \cite{Graham:2001pp,Fredenhagen:2001kw},
in agreement with the general structure of our RG flow diagram 
(see figure~\ref{fig:flowgraph}). 

As is clear from this diagram 
neither of these perturbative fixed points can be reached by a generic 
RG flow: we can only get to the unstable fixed point (II) if we first perform
the pure boundary flow to (I), followed by a pure bulk flow, and we
can only get to (III) via (II). However, we can read off the actual end-point 
of a generic flow from this picture: it is the end-point of the pure boundary 
flow starting from (II), but flowing in the opposite direction to (III). The resulting 
fixed point is therefore the non-perturbative fixed point of a certain boundary
perturbation of (II). At least in some specific cases this fixed point
has been identified before using TCSA and TBA techniques 
\cite{Chim:1995kf,Lesage,Ahn:1998xm}; applying these results
we can therefore make a prediction for the actual fixed point (IV)
of our RG flow at least for some restricted set of initial boundary conditions. 

In order to determine the corresponding fixed point for an arbitrary initial
boundary condition we finally use techniques from the perturbation theory of 
topological defects. In particular, we can identify the RG fixed point of a 
certain topological defect from the above boundary flow results (that we 
know for a restricted set of initial conditions). We can then use the RG flow
behaviour of this topological defect in order to make a prediction for the
ultimate (IV) fixed point of an arbitrary initial boundary condition. The resulting
prediction --- see eq.\ (\ref{answf}) --- is the main result of our paper.  As we shall
demonstrate it satisfies a number of consistency conditions; in particular,
the flows for different initial boundary conditions actually 
organise themselves into long chains (see figure~\ref{fig:flowchain}), 
some of whose individual flows were known before. 
In addition, we are able to confirm this prediction, for a particular
class of boundary conditions (namely those near the middle of the Kac
table), by a direct perturbative calculation --- see section~4.4. As a
final check of our prediction, we extrapolate it to small values of
$m$, and show that our results are consistent with the
numerical results of~\cite{Pearce:2000dv,Pearce:2003km}.  
\medskip

The paper is organised as follows. The coupled RG equations are worked
out in section~2.1. In section~2.2 we identify the various perturbative fixed
points and study the structure of the RG flow diagram. To support the 
identification of the fixed points we then perform a detailed analysis of the 
perturbed  $g$-function in section~3. In section~4 we  combine these
results with insights from non-perturbative flows and constraints coming 
from the action of defects to give a complete picture of the flow diagram, 
including also the actual (non-perturbative) fixed point of the original flow
(for a generic initial condition). We also discuss there various consistency 
checks which our analysis satisfies.
Finally, we compare our findings with the numerical study of 
\cite{Pearce:2000dv,Pearce:2003km} in section~5. There are three
appendices, where we collect the leading behaviour of the various OPE
coefficients (appendix~A), give explicit formulae for the bulk and boundary 
correlation functions on the upper half-plane and the disc (appendix~B), 
and analyse the explicit solutions to the RG equations (appendix~C).

\section{RG equations for minimal models}

Let us begin by reviewing our conventions. We shall consider the unitary
minimal models with central charge
\begin{equation}\label{c}
c_m=1-\frac{6}{m(m+1)}\,,\qquad {m=3,\,4,\,5,\ldots\ .}
\end{equation}
More specifically, we shall always work with the diagonal (charge conjugation) modular
invariant theory, for which the left- and the right-moving representations
are identical. For $c=c_m$, the allowed irreducible highest weight representations 
of the Virasoro algebra have highest weight 
\begin{equation}\label{hw}
h_{(r,s)} = \frac{\bigl( (m+1)r-ms \bigr)^2-1}{4m (m+1)}\ ,
\end{equation}
where  $1\leq r\leq m-1$ and $1\leq s\leq m$, and we
have the identifications  $(r,s) \cong (m-r,m+1-s)$. For the charge
conjugation theory, the conformal primary fields $\phi_i$ are labelled by
$i\equiv(r,s)$, and the conformal dimension of $\phi_i$ is 
$\Delta_i = 2 h_i$. 

We shall be interested in the conformal field theory defined on a 
Riemann surface with boundary, more specifically the 
upper half plane (or equivalently the disc). In order to characterise 
the theory on a surface with boundary we also have to specify 
the boundary conditions for the various fields. For the diagonal modular 
invariant theories, the possible conformal boundary conditions 
are also parameterised by the highest weight representations of the Virasoro
algebra. Thus the most general conformal boundary condition is a superposition
of boundary conditions associated to the irreducible representations 
labelled by $(a_1,a_2)$, where $a_1$ and $a_2$ have the same ranges and 
identifications as $r$ and $s$ in (\ref{hw}). It is sometimes convenient to 
describe the boundary condition in terms of the associated boundary state; 
for the boundary condition ${\bf a} = (a_1,a_2)$ the boundary state is given by the
Cardy formula \cite{Cardy:1989ir}
\begin{equation}\label{brane}
\braneket{{\bf a}}\equiv\braneket{a_1,a_2}=\sum_{(r,s)}
\frac{S_{(a_1,a_2)}^{\quad(r,s)}}%
{\sqrt{S_{(1,1)}^{\;(r,s)}}}\,\ishiket{r,s}\ .
\end{equation}
Here $\ishiket{r,s}$ denotes the Ishibashi state \cite{Ishibashi:1988kg} in the 
$(r,s)$ sector, and $S_{(a_1,a_2)}^{\quad(r,s)}$ is the  modular $S$-matrix, whose
entries are explicitly given as 
\begin{equation}\label{S-matrix}
S_{(a_1,a_2)}^{\quad(r,s)}=\sqrt{\frac{8}{m (m+1)}}(-1)^{1+a_1s+a_2r}\,
\sin\left(\tfrac{m+1}{m}\pi a_1r\right)
\sin\left(\tfrac{m}{m+1}\pi a_2s\right)\ .
\end{equation}
In the presence of a boundary there are also excitations located
at the boundary. These are described by boundary fields $\psi_i$,
and they are again characterised by unitary representations of the Virasoro algebra,
$i\equiv (r,s)$. The conformal dimension of the boundary field 
$\psi_i$ is given by the weight $h_i$ as in (\ref{hw}). 
On the boundary condition ${\bf a}$, the possible boundary fields 
are those that appear in the fusion rules of ${\bf a}$ with itself.  

In the following section we shall work on the upper half-plane, where we denote
the operator product expansions (OPEs) of the bulk and boundary fields as 
\begin{eqnarray}\label{OPEs}
\phi_i(z_1,\bar{z}_1)\phi_j(z_2,\bar{z}_2)
&=&\sum_{k}\tri{C}{i}{j}{k}\phi_k(z_2,\bar{z}_2)|z_1-z_2|%
^{\Delta_k-\Delta_i-\Delta_j}+\cdots\,,\nonumber\\
\phi_i(z=x+iy,\bar{z})&=&\sum_kB_i{}^k\psi_k(x)(2y)^{h_k-\Delta_i}
+\cdots\,,\\
\psi_i(x)\psi_j(y)&=&\sum_k\tri{D}{i}{j}{k}\psi_k(y)
(x-y)^{h_k-h_i-h_j} + \cdots \qquad(x>y)\,.\nonumber
\end{eqnarray}

\subsection{Bulk perturbation}

We are interested in the perturbation of the conformal field theory 
by the least relevant bulk field $\phi_{(1,3)}$ of conformal weight 
$h\equiv h_{(1,3)}=\tfrac{m-1}{m+1}$.  As is well known, this perturbation
induces an RG flow that drives the minimal model $c_m$ to the one
corresponding to $c_{m-1}$ \cite{Zamolodchikov:1987ti,Cardy:1989da}. 
In this paper we want to study what happens if we consider this 
perturbation in the presence of a boundary. It is clear that
the presence of the boundary will not affect the flow in the bulk, {\it i.e.}\ that
we are still ending up with the minimal model corresponding to $c_{m-1}$. 
However, it is not so clear what happens to the boundary condition
$(a_1,a_2)$ of the $c_m$ theory under the RG flow; this is the question we
want to address in the following.

In the presence of a boundary, the RG flow will also switch on boundary fields, and
we should therefore consider the general perturbation
\begin{equation}\label{deltaS}
\delta S=\sum_k\lambda_k\, \epsilon^{\Delta_k-2}\int d^2z\,
\phi_k(z,\bar{z})+\sum_l \mu_l \, \epsilon^{h_l-1}\int dx\,
\psi_l(x)\ .
\end{equation} 
Here the $\lambda_k$ and the $\mu_l$ are (small) dimensionless
coupling constants, and $\epsilon$ is an ultraviolet 
cut-off. As has been studied before, the combined RG equations are 
\cite{Fredenhagen:2006dn, Gaberdiel:2008fn}
\begin{eqnarray}
\dot{\lambda}_k&=&(2-\Delta_k)\lambda_k+\sum_{ij}
\pi\tri{C}{i}{j}{k}\lambda_i\lambda_j
+ {\cal O}(\lambda^3) \ ,\\
\dot{\mu}_l&=&(1-h_l)\mu_l+\sum_i\tfrac{1}{2}B_i{}^l \lambda_i
+\sum_{ij}\tri{E}{i}{j}{l}\lambda_i\mu_j+\sum_{ij}
\tri{D}{i}{j}{l}\mu_i \mu_j + 
\sum_{ij} F^l_{ij} \lambda^i\, \lambda^j +  {\cal O}(\mu^3,\lambda^2 \mu, \lambda \mu^2) \ ,
\nonumber
\end{eqnarray}
where $\dot{\lambda}$ stands for $d\lambda/d\log\epsilon$,
and similarly for $\dot{\mu}$. In the OPE scheme that we shall 
consider in the following,\footnote{To leading order in $1/m$, the calculation
is actually the same as in a minimal subtraction scheme that is more 
convenient for the calculation of the perturbed $g$-function.} the coefficients $\tri{C}{i}{j}{k}$, 
$B_i{}^l$ and $\tri{D}{i}{j}{l}$ are those from the OPEs (\ref{OPEs}).
Furthermore, the coefficient $\tri{E}{i}{j}{l}$ can be calculated
in terms of an integral of a four point function  \cite{Gaberdiel:2008fn}
(see below for more details), while the coefficient 
$F^l_{ij}$ comes from the correlation function of two bulk fields and
one boundary field. As will be explained below, this contribution
is only subleading.

Note that we have written out explicitly all terms that are of degree less or
equal to two in $\lambda$ and $\mu$. We are interested in a perturbative
fixed point for $\lambda$ and $\mu$, for which both are of 
order $1/m$. The terms we have spelled out are therefore
all the terms that contribute up to order $1/m^2$. In particular, it is important
that we also include the $\lambda\mu$ term for the $\mu$ RG equation
since it is of the same order as the $\mu^2$ term. (In fact, if one leaves it out, 
one does not find any perturbative fixed point for $\mu$.) 

The perturbation by the $(1,3)$ bulk field is particularly tractable since the 
successive OPEs of $\phi_{(1,3)}$ with itself only contain one 
relevant field apart from the identity, namely $\phi_{(1,3)}$ itself. As a 
consequence we can restrict the RG equation to $\lambda\equiv \lambda_{(1,3)}$, 
and the bulk equation is therefore
\begin{equation}
\dot{\lambda}=(2-2h)\lambda+
\pi\tri{C}{(1,3)}{(1,3)}{(1,3)}\lambda^2+\cdots\ .
\end{equation}
This leads to the well-known fixed point \cite{Zamolodchikov:1987ti,Cardy:1989da}
$\lambda_* = -\frac{1}{\pi m} + {\cal O}(m^{-2})$,
where we have used the leading behaviour of the OPE coefficients as
given in appendix~A. 

In the presence of a boundary ${\bf a}=(a_1,a_2)$, the bulk perturbation 
also induces boundary perturbations, and the only relevant boundary field
(apart from the identity) that is switched on is the boundary $\psi_{(1,3)}$ field.
Furthermore, successive OPEs of $\psi_{(1,3)}$ generate, apart from the identity
field, only one relevant field, namely $\psi_{(1,3)}$ itself. It is therefore again consistent
to restrict our attention to this field.\footnote{Obviously, this only makes sense if 
$\psi_{(1,3)}$ appears in the boundary spectrum of the boundary condition ${\bf a}=(a_1,a_2)$; 
this is the case provided that $1<a_2<m$.} Writing $\mu=\mu_{(1,3)}$, the RG equations are 
\begin{equation}\label{muRG}
\dot{\mu}=(1-h)\mu+\tfrac{1}{2}B_{(1,3)}^{\;(1,3)}\,\lambda+
\tri{E}{(1,3)}{(1,3)}{(1,3)}\,\lambda\mu+\tri{D}{(1,3)}{(1,3)}%
{(1,3)}\,\mu^2 + F \lambda^2 + \hbox{higher order}\  .
\end{equation}
The leading behaviour of the coefficients $B$ and $D$ \cite{Runkel:1998pm} are 
given in appendix~A.1, while $E$ can be calculated following 
\cite{Gaberdiel:2008fn} as 
\begin{eqnarray}\label{E_raw}
\tri{E}{(1,3)}{(1,3)}{(1,3)}&=&\lim_{\epsilon\rightarrow0}
\frac{1}{2}\int dx\,\theta(\tfrac{1}{\epsilon}-|x|)%
\bigg[ \frac{1}{\cor{\id} \tri{D}{(1,3)}{(1,3)}{(1,1)}}\,
\cor{\phi_{(1,3)}(x+\tfrac{i}{2},x-\tfrac{i}{2})%
\psi_{(1,3)}(0)\psi_{(1,3)}(\infty)} \nonumber \\
&&\phantom{\lim_{\epsilon\rightarrow0}} \qquad \qquad\qquad \qquad 
\quad-B_{(1,3)}^{\;(1,1)}\tri{D}{(1,1)}{(1,3)}{(1,3)}
-B_{(1,3)}^{\;(1,3)}\tri{D}{(1,3)}{(1,3)}{(1,3)}
\frac{\theta(|x|-1)}{|x|^h}\bigg]\ . 
\end{eqnarray}
The correlator in the first line of this expression is
a chiral four-point function which is in principle determined by the differential equation
that comes from the null vector descendant of the highest weight state with $h=h_{(1,3)}$. 
Unfortunately, this differential equation does not seem to have a simple solution. However, 
we are only interested in the chiral four point function to leading order in $1/m$, and
in this limit we find
\begin{equation}\label{correlator}
\frac{1}{\cor{\id} \tri{D}{(1,3)}{(1,3)}{(1,1)}}\, 
\cor{\phi_{(1,3)}(z,\bar{z})%
\psi_{(1,3)}(0)\psi_{(1,3)}(\infty)}=
\frac{\frac{1}{2}(z-\bar{z})^4 - \frac{3}{2}(z^4+\bar{z}^4)}
{|z|^4(z-\bar{z})^2}+\mathcal{O}(m^{-1})\ .
\end{equation}
Some remarks on the calculation can be found in appendix~B.
Note that this function has the correct asymptotic behaviour as 
$z$ approaches the boundary away from the origin, because in this
limit it goes as $\sim \tfrac{3}{4y^2}$, which is the expected
behaviour since (see appendix~A.1)
\begin{equation}
B_{(1,3)}^{\;(1,1)}\tri{D}{(1,1)}{(1,3)}{(1,3)} = 3 + {\cal O}(m^{-1}) \ .
\end{equation}
The first term in the second line in (\ref{E_raw}) subtracts precisely this
leading term. On the other hand, the other channels do not contribute 
to leading order in $1/m$ since (see again appendix~A.1)
\begin{equation}
B_{(1,3)}^{\;(1,3)}\tri{D}{(1,3)}{(1,3)}{(1,3)} = {\cal O}(m^{-1}) \  , \qquad
B_{(1,3)}^{\;(1,5)}\tri{D}{(1,5)}{(1,3)}{(1,3)} = {\cal O}(m^{-2}) \ .
\end{equation}
In particular, the integral thus converges, and we find explicitly the rather simple result
\begin{equation}\label{E}
\tri{E}{(1,3)}{(1,3)}{(1,3)}= \frac{1}{2} 
\int_{-\infty}^{\infty} dx \left[ \frac{48x^4-72x^2-5}{(1+4x^2)^2} - 3 \right] 
+ {\cal O}(m^{-1}) =  -4\pi + {\cal O}(m^{-1}) \ .
\end{equation}
Note that the leading order result (in $1/m$) is finite, and apparently
scheme independent. This ties in with the general observation of 
\cite{Gaberdiel:2008fn} that $E$ is universal provided that the resonance 
condition is satisfied, which is here the case to leading order in $1/m$.  

Finally, the term proportional to $F$ is subleading relative to these terms
since it arises from the correlation function of two bulk fields and one
boundary field. This correlation function is of the asymptotic form
\begin{equation}\label{first-lambda2-term}
\frac{1}{\cor{\id}\tri{D}{(1,3)}{(1,3)}{(1,1)}}
\cor{\phi_{(1,3)}(z_1,\bar{z}_1)\phi_{(1,3)}(z_2,\bar{z}_2)\psi_{(1,3)}(\infty)}\sim
\sum_{\mathbf{i},\mathbf{j}}B_{(1,3)}{}^{\!\mathbf{i}}B_{(1,3)}{}^{\!\mathbf{j}}
D_{\mathbf{i}\mathbf{j}}{}^{\!(1,3)}
f_{\mathbf{i}\mathbf{j}}(z_1,\bar{z}_1,z_2,\bar{z}_2)\,,
\end{equation}
where the $f_{\mathbf{ij}}(z_1,\bar{z}_1,z_2,\bar{z}_2)$ are some functions that give
the asymptotic dependence on the insertion points, and 
\mbox{$\mathbf{i},\,\mathbf{j}=(1,1),\,(1,3),\,(1,5)$}. With the OPE coefficients
from appendix~A.1, one can see that the contributions from all channels are
at most of order $1/m$. The total contribution of the $F\lambda^2$ term 
is therefore subleading relative to the other terms. 

\subsection{Analysis of fixed points}

Putting everything together, we thus have the coupled RG equations
\begin{eqnarray}
\dot{\lambda} & = & \frac{4}{m} \lambda + 4 \pi \lambda^2 +{\cal O}(\lambda^3) \nonumber \\
\dot{\mu}& = & \frac{2}{m} \mu + \frac{2\pi \alpha}{m} \lambda - 4 \pi \lambda \mu - 
\frac{4}{\alpha} \mu^2 + {\cal O}(\lambda^2 \mu, \lambda \mu^2, \mu^3) \ ,
\label{coupledRGeqs}
\end{eqnarray}
where
\begin{equation}
\alpha = \left\{ \begin{array}{ll}
(a_1 - a_2) (a_2+1) & \qquad a_2 > a_1 \vspace*{0.3cm} \\
\displaystyle{\frac{(a_2^2-1)}{m}} & \qquad a_2 = a_1 \vspace*{0.2cm} \\
(a_1 - a_2) (a_2-1) & \qquad a_1> a_2>1 \ .
\end{array} \right.
\end{equation}
These equations hold provided that $a_2>1$. Otherwise, the boundary theory
does not contain the relevant $\psi_{(1,3)}$ field, and there is therefore no equation
for $\dot\mu$. 

Apart from the trivial fixed point $(\lambda=\mu=0)$, these equations have
the following three fixed points (for $a_2>1$):
\begin{list}{(\Roman{enumi})}{\usecounter{enumi}}
\item The fixed point at  
\begin{equation}
\lambda_* = 0 \ , \qquad  \mu_* = \frac{\alpha}{2m} \ .
\end{equation}
This is simply the perturbative fixed point in the pure boundary analysis of 
\cite{Recknagel:2000ri} (see also \cite{Graham:2001pp}). 
As was explained there, it describes the flow
\begin{equation}\label{RRS}
(a_1, a_2)_m \quad  \longrightarrow  \quad
\bigoplus_{l=1}^{\min(a_1,a_2)} (a_1+a_2+1-2l,1)_m \ .
\end{equation}
The end-point of the flow is a boundary condition in the $m^{\rm th}$ theory. 
\item The fixed point at 
\begin{equation}\label{mu*}
\lambda_* = - \frac{1}{\pi m} \ , \qquad \mu_* = \frac{\alpha}{2m}  \ .
\end{equation}
The interpretation of this fixed point will be determined in detail in section~3, 
where we will show that it describes the superposition of boundary conditions
in the $(m-1)^{\rm st}$ theory 
\begin{equation}\label{II}
(a_1, a_2)_m \quad \longrightarrow  
\quad \bigoplus_{l=1}^{\min(a_1,a_2)} (1,a_1+a_2+1-2l)_{m-1} \ .
\end{equation}
\item The fixed point at 
\begin{equation}
\lambda_* = - \frac{1}{\pi m} \ , \qquad \mu_* = \frac{\alpha}{m}  \ .
\end{equation}
As we will also show in section~3, this fixed point describes the end-point of a
perturbative boundary flow in the $(m-1)^{\rm st}$ theory, starting from the 
boundary condition at (II). The end-point describes the boundary condition
\begin{equation}\label{IItoIII}    
(a_1,a_2)_m \qquad \stackrel{\hbox{\footnotesize via (II)}}{\longrightarrow} 
\qquad (a_2,a_1)_{m-1} \ ,
\end{equation}
in agreement with a boundary flow of \cite{Graham:2001pp}, where it
was observed that by turning on all $\psi_{(1,3)}$ boundary condition
changing fields one can find the following two perturbative flows,
\begin{equation}\label{kevinsflows}
 \bigoplus_{l=1}^{\min(a_1,a_2)} (1,a_1+a_2+1-2l)_{m-1}
 \quad \longrightarrow \quad \left\{\begin{array}{l}
 (a_{2},a_{1})_{m-1}\\
 (a_{1},a_{2})_{m-1}
\end{array} \right. 
\end{equation} 
(see eqs (5.29) and (5.30) in~\cite{Graham:2001pp}).
\end{list}

The above analysis applies to the case when both labels $a_1$ and $a_2$ are 
small, and $a_2>1$. If $a_2=1$, 
on the other hand,
there is just the RG flow for $\lambda$, whose fixed point is 
$\lambda_*=-\frac{1}{\pi m}$.  As will also be explained in section~3, it
describes the flow
\begin{equation}\label{a2=1}
(a_1,1)_m \quad \longrightarrow \quad (1,a_1)_{m-1} \ .
\end{equation}

It is easy to see from the flow diagram in the
introduction (see figure~\ref{fig:flowgraph}) that for 
$a_2>1$ the actual flow cannot directly reach the fixed 
points (II)  or (III).  In fact, starting from $\lambda=\mu=0$ we 
do not get to the fixed 
point (II), unless we first perform the pure boundary flow leading to (I), followed 
by the pure bulk flow (\ref{a2=1}). To reach the fixed point (III) we first have to go
to (II) via (I), and then have to switch on a pure boundary perturbation at (II). Indeed,
the fixed point (II) is again unstable since the 
boundary condition (\ref{II}) has at least one relevant $\psi_{(1,3)}$ field in its 
spectrum. This can also be seen by expanding the RG equation around the 
fixed point (II) by setting $\lambda=\lambda_{*}$ and $\mu =\alpha/2m + \tilde{\mu}$,
\begin{equation}\label{2.25}
\dot{\tilde{\mu}} \Big|_{\lambda =\lambda_{*}} =
\frac{2}{m}\tilde{\mu} -\frac{4}{\alpha}\tilde{\mu}^{2} + \dotsb \ .
\end{equation}
The coefficient of the term linear in $\tilde{\mu}$ allows us to read off 
the conformal weight of the boundary field to which $\tilde\mu$ couples,
and one finds indeed $h_{\tilde\mu} = h_{(1,3)} = 1-\frac{2}{m}+\dotsb$. 

Given the various different kinds of flows we can consider, our resulting picture
will have to satisfy a number of consistency conditions. These will be discussed in
section~4, where we shall also analyse the actual non-perturbative fixed point (IV)
to which a generic initial configuration will flow. Before we discuss these issues, let us
first analyse the perturbed $g$-function in order to identify the different
perturbative fixed points.

\section{Analysis of the perturbed $g$-function}
\setcounter{equation}{0}

In order to corroborate our above claims about the perturbative fixed points, 
we shall now calculate the perturbed boundary entropy, 
as was done for the case of the pure boundary perturbation in \cite{Recknagel:2000ri}. 
In order to be able to compare with their results, we shall now work on the disc. 

Recall that the boundary entropy $g({\bf a})$ of a boundary condition ${\bf a}$ is defined 
to be the correctly normalised one-point function of the identity operator in the presence 
of the boundary  condition ${\bf a}$ \cite{Affleck:1991tk} 
\begin{equation}\label{gfunction}
g^{(m)}({\bf a})=
\frac{S_{\bf a}^{\;\id}}{{\sqrt{S_{\id}^{\;\id}}}}=
\left(\frac{8}{m (m+1)}\right)^{\frac{1}{4}}
\frac{\sin\frac{\pi a_1}{m} \,\sin\frac{\pi a_2}{m+1}}
{(\sin\frac{\pi}{m}\, \sin\frac{\pi}{m+1})^{\frac{1}{2}}
} \ .
\end{equation}
It was conjectured in \cite{Affleck:1991tk} and perturbatively verified
in \cite{Affleck:1992ng,Recknagel:2000ri} that this quantity decreases under 
pure boundary RG flows. Obviously, the same need not be true in
our context, since our perturbation also changes the bulk theory, and 
we are therefore comparing one-point functions in different bulk models 
\cite{Dorey:1999cj} (see also \cite{Green:2007wr}). 

We should thus not expect to be able to say much about the
overall $g$-functions. On the other hand, the relative $g$-functions
should continue to have a well-defined meaning (see also the discussion in 
\cite{Dorey:1999cj}). Furthermore, we expect that 
the boundary condition ${\bf a}=(1,1)$ should flow to itself, since the boundary
spectrum of the $(1,1)$ boundary does not contain any relevant fields
(apart from the identity), and since $(1,1)$ is the boundary condition with the smallest
$g$-function. Thus it is natural to consider the relative boundary entropy with 
respect to ${\bf a}=(1,1)$, 
\begin{equation}
\hat{g}^{(m)}({\bf a}) = \frac{g^{(m)}({\bf a})}{g^{(m)}(1,1)} =
\frac{\sin\frac{\pi a_1}{m} \,\sin\frac{\pi a_2}{m+1}}
{\sin\frac{\pi}{m}\, \sin\frac{\pi}{m+1}} \ .
\end{equation}

\noindent For the following it will be important to consider the
asymptotic expansion of the $g$-function for large $m$ with fixed
boundary labels $a_{1},a_{2}$,
\begin{equation}\label{ghasymptotics}
\hat{g}^{(m)}({\bf a})=a_1a_2\left(1-\frac{\pi^2}{6}\bigg(
a_1^2+a_2^2-2\bigg)\frac{1}{m^2}+\frac{\pi^2}{3}\bigg(
a_2^2-1\bigg)\frac{1}{m^3}+\mathcal{O}(m^{-4})\right)\ .
\end{equation}
Our aim is therefore to calculate the perturbed $g$-function of the
boundary condition ${\bf a}$ up to order $1/m^3$, and to deduce from
it the perturbed value of the relative $g$-function (up to this
order). This should then be identified with the relative $g$-function
(in the $c_{m-1}$ theory) of the boundary condition to which ${\bf a}$
flows to. Let us first deal with the case where $a_2>1$, so that the
original boundary condition has a relevant boundary field in its
spectrum.

\subsection{The analysis for $a_2>1$}

Since we are interested in ratios of $g$-functions, it is convenient to consider the
logarithm of the perturbed $g$-function
\begin{eqnarray}\label{perturbedg}
\log \cor{e^{\delta S}}_{\bf a} 
&= &\log \cor{\id}_{\bf a}+ \frac{1}{\cor{\id}_{\bf a}}\bigg(
\frac{1}{2}\mu_*^2\epsilon^{2h-2}\int dw_1dw_2\cor{\psi(w_1)\psi(w_2)}_{\bf a}^{c} \\
&&\qquad\qquad\qquad\quad
+\frac{1}{6}\mu_*^3\epsilon^{3h-3}\int dw_1dw_2dw_3\cor{\psi(w_1)
\psi(w_2)\psi(w_3)}_{\bf a}^{c} \nonumber\\
& & \qquad \qquad\qquad \quad
+ \lambda_*\epsilon^{2h-2}\int d^2u\cor{\phi(u,\bar{u})}_{\bf a}^{c} 
+ \lambda_*\mu_*\epsilon^{3h-3}\int d^2u\, dw
\cor{\phi(u,\bar{u})\psi(w)}_{\bf a}^{c} \nonumber\\ 
&&\qquad\qquad\qquad\quad
+\frac{1}{2}\lambda_*\mu_*^2\epsilon^{4h-4}\int d^2u\,dw_1dw_2
\cor{\phi(u,\bar{u})\psi(w_1)\psi(w_2)}_{\bf a}^{c}+\cdots \bigg)\ .\nonumber 
\end{eqnarray}
Here the suffix $c$ at the correlators indicates that we are only considering
the connected components; for the terms that we have written explicitly above,
this only makes a difference for the last contribution. 

In order to identify the fixed points we need to evaluate the perturbed $g$-function 
up to order $1/m^3$. Since both $\lambda_*$ and $\mu_*$ are of 
order $1/m$, we only need to consider terms that are at most of cubic order 
in these coupling constants.\footnote{To leading order in $1/m$ the bare and the 
renormalised coupling constants agree, and we we can therefore directly use the 
above fixed points (for the renormalised coupling constants) here.} 
 We have written out explicitly all such terms except those that are proportional to 
$\lambda^2$. The reason for this is that to order $1/m^3$, they turn out not to 
depend on the boundary labels $(a_1,a_2)$, and therefore will not contribute 
to the relative entropy at order $1/m^3$. This can be seen from considering the 
possible asymptotics of the respective correlators. For the correlator of two bulk 
fields in the $\lambda^2$ term there are the asymptotic channels
${\bf i,j}=(1,1),\,(1,3),\,(1,5)$, and we have
\begin{equation}
\cor{\phi(u_1,\bar{u}_1)\phi(u_1,\bar{u}_1)}_{\bf a}\sim%
\sum_{\bf i,j}B_{(1,3)}{}^{\!{\bf i}}\,B_{(1,3)}{}^{\!{\bf j}}\,
\tri{D}{{\bf i}}{{\bf j}}{(1,1)}%
\cor{\id}_{\bf a}f_{{}\bf ij}(u_1,\bar{u}_1,u_2,\bar{u}_2)\ ,
\end{equation}
where the $f_{{\bf ij}}$ are some functions that give the asymptotic dependence on 
the insertion points, compare \eqref{first-lambda2-term}. 
With the OPE constants from appendix~A.1 one can see that
all channels contribute boundary-dependent terms only at order $1/m^2$,
so that the whole contribution of the $\lambda^2$ term will be of order $1/m^4$. 
For the terms with coupling constants $\lambda^2\mu$ and $\lambda^3$, a similar analysis
shows that their contributions will only affect the order $1/m^4$ as well.

The above integrals are not well defined, and we need to introduce a regularisation 
scheme to make sense of them. In each case the leading term in $1/m$ that is 
dependent on the boundary labels $(a_1,a_2)$ turns out to be of order $1/m^3$,
and thus we are effectively working to leading order in $1/m$. In particular, we can 
therefore take the large $m$ expansion of the integrand before we do the integral. 
This integral will be regularised by some cut-off scheme; in particular, we shall
introduce a cut-off $\epsilon$ to separate the boundary fields from one another, and a 
cut-off $\xi$ to restrict the radial bulk integration from $0\leq r \leq 1-\xi$. We shall
then discard the terms proportional to $\epsilon^{-1}$, as they describe non-universal
terms (that have the wrong scaling behaviour). We shall also impose a similar
procedure for the terms that are singular in $\xi\rightarrow 0$ which we shall describe
in more detail below. 

We should note that this cut-off regularisation scheme is obviously
not the same as the scheme with which the RG equations of section~2 were derived. 
However, to leading order in $1/m$ the quantities we calculate should be universal,
and thus this distinction should not play a role; this expectation will be borne out
by our results. We have also checked this explicitly for some of the terms; the 
advantage of the cut-off scheme we are using here is that the calculations are much
simpler since we do not need to know the integrand for arbitrary $m$ (but only
in the $m\rightarrow \infty$ limit). 

We shall now discuss the various terms in turn. The first two terms are the 
pure boundary integrals that were already considered in \cite{Recknagel:2000ri}. 
The boundary contribution proportional to $\mu_*^2$ involves the integral
\begin{eqnarray}
I_1&=&\int d\theta_1 d\theta_2\,\left|2\sin\frac{\theta_1-\theta_2}{2}
\right|^{-2h} = \frac{\pi}{2} \int_\epsilon^{2\pi-\epsilon} d\theta\left(\sin^{-2}
\frac{\theta}{2}+\frac{4\log\left(2\sin\frac{\theta}{2}\right)}{\sin^2%
\frac{\theta}{2}}\frac{1}{m}+\mathcal{O}(m^{-2})\right)\nonumber\\
&=&2\pi\bigg(\cot\frac{\epsilon}{2}+\frac{2}{m}\left(\epsilon-\pi+
\cot\frac{\epsilon}{2}\,\left(1+\log(2\sin\frac{\epsilon}{2})\right)\right)\bigg)
+\mathcal{O}(m^{-2})\nonumber\\
&=&\frac{4\pi}{\epsilon}+\left(\frac{16\pi(1+\log\epsilon)}{\epsilon}-
4\pi^2\right)\frac{1}{m}+\mathcal{O}(\epsilon,m^{-2})\ .
\end{eqnarray}
The first nontrivial contribution of the integral is hence of order $1/m$, and
we obtain the contribution (dropping the non-universal term proportional to
$1/\epsilon$)
\begin{eqnarray}
\frac{1}{2}\mu_*^2\epsilon^{2h-2}\int dw_1dw_2\cor{\psi(w_1)\psi(w_2)}_{\bf a}^{c}
& = &  -\pi^2\, \frac{2}{m}\, \tri{D}{(1,3)}{(1,3)}{(1,1)}\cor{\id}_{\bf a} \mu_*^2 +
{\cal O}(m^{-4})\nonumber \\
& = & - \frac{4\pi^2}{m}\, 
\left( \frac{\mu_*}{\alpha}\right)^2\, \cor{\id}_{\bf a} (a_2^2-1)
+ \mathcal{O}(m^{-4}) \ ,
\end{eqnarray}
where the correction term is of order $m^{-4}$ for all fixed points of interest. 
The boundary contribution depending on $\mu_*^3$ involves the integral
\begin{eqnarray}
I_2&=&\int d\theta_1d\theta_2d\theta_3\,\left|8\sin\frac{\theta_{12}}{2}\,
\sin\frac{\theta_{23}}{2}\,\sin\frac{\theta_{31}}{2}\right|^{-h}\nonumber\\
&=&4\pi\int_{\theta_1>\theta_2} d\theta_1d\theta_2\,\left(8\sin\frac{\theta_{12}}{2}\,
\sin\frac{\theta_{1}}{2}\,\sin\frac{\theta_{2}}{2}\right)^{-1}+\mathcal{O}(m^{-1})\ .
\end{eqnarray}
If we introduce cut-offs only where necessary, so that the integral
does not diverge, we find
\begin{eqnarray}
I_2&=&8\pi\cot\frac{\epsilon}{2}-4\pi^2+\mathcal{O}(m^{-1} ) = 
\frac{16\pi}{\epsilon}-4\pi^2+\mathcal{O}(\epsilon,m^{-1})\,.
\end{eqnarray}
Again dropping the non-universal $1/\epsilon$ term, we get the contribution
\begin{eqnarray}
\frac{1}{6}\mu_*^3\epsilon^{3h-3}\int dw_1dw_2dw_3\cor{\psi(w_1)
\psi(w_2)\psi(w_3)}_{\bf a}^{c} & = & 
\frac{2\pi^2}{3}\tri{D}{(1,3)}{(1,3)}{(1,3)}
\tri{D}{(1,3)}{(1,3)}{(1,1)}\cor{\id}_{\bf a} \mu_*^3 + {\cal O}(m^{-4}) \nonumber \\
& = & \frac{16 \pi^2}{3}\ 
\left(\frac{\mu_*}{\alpha}\right)^3\, \cor{\id}_{\bf a}\, (a_2^2-1)
+\mathcal{O}(m^{-4}) \,.
\end{eqnarray}
The contribution which depends linearly on $\lambda_*$ and is 
independent of $\mu_*$  involves the integral
\begin{eqnarray}
I_3 &=&\int dr\,r(1-r^2)^{-2h} = \int_0^{1-\xi} dr\,r(1-r^2)^{-2}+\mathcal{O}(m^{-1})\nonumber\\
&=&\frac{(1-\xi)^2}{2\xi(2-\xi)}+\mathcal{O}(m^{-1}) = 
\frac{1}{8}\,\frac{2-3\xi}{\xi}+\mathcal{O}(\xi,m^{-1})\ .
\label{I3}
\end{eqnarray}
This yields
\begin{eqnarray}\label{lambda--term}
\lambda_*\epsilon^{2h-2}\int d^2u\cor{\phi(u,\bar{u})}_{\bf a}^{c}=
\lambda_*\cor{\id}_{\bf a}\,B_{(1,3)}^{\;(1,1)}\,\left(\frac{\pi}{4}\,%
\frac{2-3\xi}{\xi}+\mathcal{O}(\xi,m^{-1})\right)\nonumber\\
=\lambda_*\cor{\id}_{\bf a}\left(-\frac{\pi^3(a_2^2-1)}{m^2}\,%
\frac{2-3\xi}{\xi}+f(m,\xi)+\mathcal{O}(\xi,m^{-1})\right)\ ,
\end{eqnarray}
where $f(m,\xi)$ is some function which does not depend on the boundary
labels and has the large-$m$ asymptotic behaviour
\begin{equation}\label{3.10}
f(m,\xi)=\frac{3\pi}{4}\,\frac{2-3\xi}{\xi}+\mathcal{O}(\xi,m^{-1})\ .
\end{equation}
The contribution that depends linearly on $\lambda_*$ and $\mu_*$ involves the 
integral
\begin{eqnarray}
I_4&=&\int drd\theta_1 d\theta_2\,r((1-r^2)(1-2r\cos(\theta_1-\theta_2)+r^2))^{-h} \nonumber \\
&=&\int_0^{1-\xi} dr\,\frac{4\pi^2r}{(1-r^2)^2}+\mathcal{O}(m^{-1})
=\frac{\pi^2}{2}\frac{2-3\xi}{\xi}+\mathcal{O}(\xi,m^{-1})\ ,
\end{eqnarray}
leading to 
\begin{eqnarray}
\lambda_*\mu_*\epsilon^{3h-3}\int d^2u\, dw
\cor{\phi(u,\bar{u})\psi(w)}_{\bf a}^{c}&=&
\lambda_*\mu_*\cor{\id}_{\bf a}B_{(1,3)}^{\;(1,3)}\,\tri{D}{(1,3)}{(1,3)}{(1,1)}\,
\left(\frac{\pi^2}{2}\frac{2-3\xi}{\xi}+\mathcal{O}(\xi,m^{-1})\right) \nonumber \\
& = & \lambda_* \left(\frac{\mu_*}{\alpha}\right)\, 
\cor{\id}_{\bf a} \frac{4\pi^3}{m}\, (a_2^2-1) \left(\frac{2-3\xi}{\xi}+
\mathcal{O}(\xi,m^{-1})\right)
\end{eqnarray}
Finally, the integrand for the term of order $\lambda_* \mu_*^2$ is the same as 
in the calculation of the coefficient $E$ in the RG equation of section~2. Deferring
the details of this calculation to appendix~B the resulting contribution turns out to be 
\begin{eqnarray}
&&\frac{1}{2}\lambda_*\mu_*^2\epsilon^{4h-4}\int d^2u\,dw_1dw_2
\cor{\phi(u,\bar{u})\psi(w_1)\psi(w_2)}_{\bf a}^{c}\nonumber\\
&&\quad =\frac{\pi}{3}\lambda_*\mu_*^2\cor{\id}_{\bf a}
B_{(1,3)}^{\;(1,1)}\,\tri{D}{(1,1)}{(1,3)}{(1,3)}\,\tri{D}{(1,3)}{(1,3)}{(1,1)}
\left(-2\pi^2\,\frac{2-3\xi}{\xi}+\mathcal{O}(\xi)+\cdots \right) \nonumber \\
&&\quad =- \lambda_*\left(\frac{\mu_*}{\alpha}\right)^2\, \cor{\id}_{\bf a}\,
4\pi^3 (a_2^2-1) \frac{2-3\xi}{\xi}+\mathcal{O}(\xi,m^{-4})\ ,
\label{phipsipsi-contribution}
\end{eqnarray} 
where the ellipses in the second line refer to terms of order ${\cal O}(\epsilon,m^{-1})$.
Adding all the relevant terms together we then arrive at 
\begin{eqnarray}\label{perturbedans}
\log \cor{e^{\delta S}}_{\bf a} 
&= &\log \cor{\id}_{\bf a}+ (a_2^2-1) \Bigl[
- \frac{4\pi^2}{m} \left(\frac{\mu_*}{\alpha}\right)^2 
+ \frac{16\pi^2}{3} \left(\frac{\mu_*}{\alpha}\right)^3  \nonumber \\
& & \qquad \qquad + \lambda_* \, \frac{2-3\xi}{\xi} \Bigl(
-\frac{\pi^3}{m^2} + 
\frac{4\pi^3}{m}  \left(\frac{\mu_*}{\alpha}\right) 
- 4\pi^3 \left(\frac{\mu_*}{\alpha}\right)^2 \Bigr) \Bigr] + \cdots  \ ,
\end{eqnarray}
where the ellipses either denote terms that are independent of the boundary
labels, or terms that are of order ${\cal O}(m^{-4})$ for the fixed points of interest. 
The first two coefficients (that are independent of $\lambda_*$ 
reproduce exactly what was found in \cite{Recknagel:2000ri}. Using
their results it thus follows that the fixed point (I) is indeed the perturbative
fixed point of the pure boundary perturbation. 

\subsubsection{Fixed point (II)}

At the fixed point (II), we have $\frac{\mu_*}{\alpha} = \frac{1}{2m}$, and it
follows from (\ref{perturbedans}) that the last three terms (that are all proportional
to $\lambda_*$) cancel identically. The perturbed $g$ function 
(\ref{perturbedans}) is then given by
\begin{equation}
\log g_{\lambda_*,\mu_*}^{(m)}({\bf a})= \log  g^{(m)}({\bf a})
-\frac{\pi^2}{3m^3}(a_2^2-1)+f(m)
+\mathcal{O}\bigl(m^{-4} \bigr)\ ,
\end{equation}
where $f(m)$ is a function which is at least of order 
$1/m$ and does not depend on the boundary labels.
Subtracting the corresponding expression for the boundary entropy of the
$(1,1)$ boundary condition, we thus find that the perturbed relative
entropy $\hat{g}$ equals
\begin{eqnarray}\label{right}
\hat{g}^{(m)}_{\lambda_*,\mu_*}({\bf a})&=& \hat{g}^{(m)}({\bf a})\,
\left(1-\frac{\pi^2}{3}(a_2^2-1)\frac{1}{m^3}+
\mathcal{O}(m^{-4})\right)\nonumber\\
&=&a_1a_2\left(1-\frac{\pi^2}{6}(a_1^2+a_2^2-2)\frac{1}{m^2}+
\mathcal{O}(m^{-4})\right)\ ,
\end{eqnarray}
where in the final line we have used the asymptotic expansion
(\ref{ghasymptotics}) for $\hat{g}^{(m)}({\bf a})$.

This is now to be compared with the relative entropy of the boundary
condition ${\bf b}= (b_1,b_2)$ in the $c_{m-1}$ theory, 
\begin{equation}\label{left}
\hat{g}^{(m-1)}({\bf b})=b_1b_2\left(1-\frac{\pi^2}{6}%
(b_1^2+b_2^2-2) \frac{1}{m^2}-\frac{\pi^2}{3}(b_1^2-1) \frac{1}{m^3}
+\mathcal{O}(m^{-4})\right)\ .
\end{equation}
If $a_1=1$, (\ref{left}) equals (\ref{right}) for
a single fundamental boundary condition ${\bf b}$
\begin{equation}
{\bf a}=(1,a_2)_m \quad \stackrel{{\rm (II)}}{\longrightarrow} \quad  {\bf b}=(1,a_2)_{m-1}\ .
\label{singleflow2}
\end{equation}
In the general case, we cannot solve for 
(\ref{right}) = (\ref{left}) with a single boundary condition 
${\bf b}=(b_1,b_2)$.\footnote{In \cite{Green:2007wr} a 
similar calculation was done to lower order in $1/m$, from
which the authors concluded that the flow is simply
$(a_1,a_2) \rightarrow (a_2,a_1)$. This is compatible with
the analysis to order $1/m^2$, but not to order $1/m^3$.}
As in \cite{Recknagel:2000ri} we therefore consider 
superpositions of fundamental boundary conditions
\begin{equation}
{\bf B}=\bigoplus_{l=1}^{N} {\bf b}^l\,,\qquad
{\bf b}^l=(b_1^l,b_2^l)\ .
\end{equation}
The entropy of the superposition is just the sum of the individual
entropies, and we thus get the equations 
\begin{eqnarray}
\sum_lb_1^lb_2^l& =& a_1a_2\ ,\nonumber\\
\sum_lb_1^lb_2^l \Bigl((b_1^l)^2 + (b_2^l)^2 - 2\Bigr)&=&
a_1 a_2 \Bigl( a_1^2 + a_2^2 - 2\Bigr)\ ,\\
\sum_l  b_1^lb_2^l  \Bigl((b_1^l)^2 - 1\Bigr)&=&0\ .\nonumber
\end{eqnarray}
The last equation implies that $b_1^l=1$ for all $l$, and the equations are generically 
solved by
\begin{equation}
N=\mathrm{min}(a_1,a_2)\ ,\qquad
b_1^l=1\quad\forall l\ ,\qquad
b_2^l=a_1+a_2+1-2l\ .
\end{equation}
This suggests flows of the form
\begin{equation}\label{answ}
\mathbf{a}=(a_1,a_2)_m\quad\stackrel{{\rm (II)}}{\longrightarrow} \quad
\mathbf{B}=
\sum_{l=1}^{\mathrm{min}(a_1,a_2)}(1,a_1+a_2+1-2l)_{m-1}\ .
\end{equation}
These flows have to be understood as sequences of flows (first a pure
boundary flow to (I) followed by the flow to (II)), as was discussed at the
end of section~2 (see also figure~\ref{fig:flowgraph}).  Note that the
result of~\eqref{answ} is similar to what happened in the case of a
pure boundary perturbation \cite{Recknagel:2000ri}, except that there
the end-points of the boundary flow were superpositions of boundary
conditions $(b_l,1)$, while here we have boundary conditions
$(1,b_l)$.

We should mention that the $g$ function does not allow us to determine
the resulting boundary conditions uniquely, since
\begin{equation}\label{ambiguity}
\hat{g}^{(m-1)}(b_1,b_2) = \hat{g}^{(m-1)} (b_1,m-b_2) \ .
\end{equation}
One can partially fix this ambiguity as in \cite{Recknagel:2000ri}. To zeroth order
in perturbation theory, we know that the $(r,s)$ bulk field of the $m^{\rm th}$ theory
becomes the $(s,r)$ bulk field of the $(m-1)^{\rm st}$ theory 
\cite{Zamolodchikov:1987ti}. Thus
to leading order in $1/m$ we need to have that
\begin{equation}
\left(^{\bf a}B_{(r,s)}^{\;(1,1)}\right)^{(m)}  = \left(^{\bf
B}B_{(s,r)}^{\;(1,1)} \right)^{(m-1)} \ .
\end{equation}
Since 
\begin{equation}
\left(^{(a_{1},a_{2})}B_{(r,s)}^{\;(1,1)}\right)^{(m)} = 
(-1)^{(r+s)(a_1+a_2)} \, 
\frac{\sin\left(\frac{r a_1 \pi}{m}\right) \, \sin\left(\frac{s a_2 \pi}{m+1}\right)}%
{\sin\left(\frac{a_{1} \pi}{m}\right) \, \sin\left(\frac{a_{2} \pi}{m+1}\right)} \ ,
\end{equation}
this requires, in particular, that 
\begin{equation}
(-1)^{(r+s)(a_1+a_2)} = (-1)^{(r+s)(b_1^l + b_2^l)} 
\end{equation}
for all $(r,s)$ and all $l$. This is evidently satisfied by our ansatz, but would not in 
general be true if we replaced some $(b_1^l,b_2^l)$ by $(b_1^l,m-b_2^l)$. 

\subsubsection{Fixed point (III)}

The analysis for the fixed point (III) is more complicated, since now the $\lambda_*$
term in the last line of (\ref{perturbedans}) does contribute. We therefore need to 
understand how to deal with the singular part as $\xi\rightarrow 0$. We do not 
have a fundamental understanding of how this must be done, but we shall
now propose a procedure that will lead to consistent results. In fact, the procedure
is already determined by considering the consistency of the flow of the $(a_1,1)$ 
boundary condition (see section~4, as well as the remarks below in section~3.2), 
and the fact that all the other consistency  conditions then also work out 
(that will be explained in detail in section~4) makes us confident that this is 
indeed the correct prescription.\footnote{The prescription we
propose actually has a very natural interpretation if we consider the theory on
the semi-infinite cylinder (see appendix~B.2). We thank Anatoly
Konechny for discussions on this point.} 

Since the cut-off $\xi$ bounds the radial bulk integral to the region 
$0\leq r \leq 1-\xi$, it is actually a measure of an area; in fact, the area of the
missing annulus is simply $\pi (1-(1-\xi)^2) = 2 \pi \xi (1 - \xi/2)$. In order 
to convert it to a parameter that scales as a length, we should divide it 
by a natural length, which we take to be the circumference of the inner circle, 
{\it i.e.}\ $2 \pi (1-\xi)$. We thus propose that the good cut-off parameter is 
\begin{equation}
\eta = \frac{\xi\, (1-\xi/2)}{(1-\xi)} = \xi ( 1 + \xi/2) + {\cal O}(\xi^3) \ .
\end{equation}
The expression of interest can therefore be written as 
\begin{equation}\label{etapres}
\frac{2-3\xi}{\xi} =  \frac{2}{\eta} - 2 + {\cal O}(\eta) \ .
\end{equation}
Our prescription is now that we should discard the non-universal 
$1/\eta$ pole, but keep the constant term. Using that at (III) 
$\frac{\mu_*}{\alpha} =  \frac{1}{m}$ as well as $\lambda_* = - \frac{1}{\pi m}$
then leads to 
\begin{equation}
\log g_{\lambda_*,\mu_*}^{(m)}({\bf a})= \log  g^{(m)}({\bf a})
-\frac{2 \pi^2}{3m^3}(a_2^2-1)+f(m)
+\mathcal{O}\bigl(m^{-4} \bigr)\ .
\end{equation}
The perturbed relative entropy $\hat{g}$ then equals 
\begin{eqnarray}\label{rightIII}
\hat{g}^{(m)}_{\lambda_*,\mu_*}({\bf a})&=& \hat{g}^{(m)}({\bf a})\,
\left(1-\frac{2\pi^2}{3}(a_2^2-1)\frac{1}{m^3}+
\mathcal{O}(m^{-4})\right)\nonumber\\
&=&a_1a_2\left(1-\frac{\pi^2}{6}(a_1^2+a_2^2-2)\frac{1}{m^2}
- \frac{\pi^2}{3 m^3} (a_2^2-1) + 
\mathcal{O}(m^{-4})\right)\ ,
\end{eqnarray}
where in the final line we have used the asymptotic expansion
(\ref{ghasymptotics}) for $\hat{g}^{(m)}({\bf a})$. Comparing this
with (\ref{left}) we then find that the end-point should be given by 
\begin{equation}
(a_1,a_2)_m \qquad
\stackrel{{\rm (III)}}{\longrightarrow}\qquad (a_2,a_1)_{m-1} \ .
\end{equation}
This flow has to be understood as a sequence of asymptotic flows, similar
to~\eqref{answ}.

\subsection{The analysis for $a_2=1$}

For $a_2=1$ the analysis is simpler since the boundary 
condition does not have the relevant boundary field $\psi_{(1,3)}$ in its spectrum, and 
thus we have no RG equation for $\mu$. Of the terms in (\ref{perturbedg}) hence only 
the term (\ref{lambda--term}) survives. However, since it does not depend on $a_1$, and 
since we are only interested in the ratio of the perturbed $g$-function relative to 
${\bf a}=(1,1)$ it does not contribute to the rescaled $g$ function in the $(m-1)^{\rm st}$ 
theory. Thus we conclude that the (rescaled) ${g}$-function does not change at all. 
Using eq.~(\ref{gfunction}) we find that 
\begin{equation}
\hat{g}^{(m)}(a_1,1) = \frac{\sin\frac{\pi a_1}{m}}{\sin\frac{\pi}{m}} = 
\hat{g}^{(m-1)}(1,a_1) \ .
\end{equation}
This then establishes (\ref{a2=1}). Note that this argument holds for
arbitrary values of $a_1$, not necessarily small relative to $m$.

Actually, since in this case no combined flow takes place, one may suspect 
that not only the ratio of $g$-functions is correctly reproduced by this analysis, 
but also the overall value of the $g$-function. As in the pure boundary case, 
let us consider the logarithmic change of the $g$ function. The only term
that contributes in this case is the term (\ref{lambda--term}) that contains
a contribution of order $1/m$ that is independent of the boundary 
labels. Using (\ref{3.10}) we thus have to first order 
\begin{equation}\label{bulkpartforxi1}
\log \frac{g_{\lambda_*,\mu_*}^{(m)}(a_1,1)}{g^{(m)}(a_{1},1)}=
-\frac{3}{4m}\,\frac{2-3\xi}{\xi}+\mathcal{O}(\xi,m^{-2})\,,
\end{equation}
which we must compare with the expansion of the resulting boundary condition,
\begin{equation}\label{bulkpartforxi2}
\log \frac{g^{(m-1)}(1,a_1)}{g^{(m)}(a_1,1)}=
\frac{3}{2m}+\mathcal{O}(m^{-2})\,.
\end{equation}
These two expressions then agree precisely to this order if we use
the same prescription as in (\ref{etapres}).

\section{The actual fixed point and consistency constraints}
\setcounter{equation}{0}

So far we have identified the perturbative fixed points by studying
the perturbed $g$-function. The actual fixed point (IV) of the RG flow for
a generic initial condition however appears at 
$\lambda_* = - \frac{1}{\pi m}$ and $\mu_*=+\infty$ (for $\alpha<0$ --- for
$\alpha>0$ we have instead $\mu_*=-\infty$),  as is clear from the flow
diagram (see figure~\ref{fig:flowgraph}).
We now want to combine what we have
found so far with results about non-perturbative boundary flows in
order to identify this actual (non-perturbative) fixed point. In order
to tie it down completely, we shall also use some constraints that
arise upon using defects.

First of all, one way to get to (IV) is to turn on a $\psi_{(1,3)}$ field on the 
boundary condition (II). Indeed, as we have shown above in (\ref{2.25}) the 
coupling constant $\tilde{\mu}$ couples to a $\psi_{(1,3)}$ field at the fixed point (II). 
Furthermore, the flow from (II) to (III) is the usual perturbative
$\psi_{(1,3)}$ flow, and thus the flow from (II) to (IV) must be the corresponding
non-perturbative flow (where the field $\psi_{(1,3)}$ is switched on with the opposite
sign). The situation is particularly simple if the boundary spectrum of (II) only
contains a single $\psi_{(1,3)}$ field --- then it is clear that this is the field to
which $\tilde{\mu}$ must couple. This will be the case provided that we begin
with a boundary condition of type $(1,a_2)_m$, since then the perturbative fixed point
(II) consists of a single boundary condition $(1,a_2)_{m-1}$ that has a single 
$\psi_{(1,3)}$ boundary field in its spectrum. The fixed point (III)
is then the end-point $(a_2,1)_{m-1}$ of the 
perturbative boundary flow (provided that $a_2$ is not too large)
\cite{Recknagel:2000ri}.  

In this case we can determine the end-point of the non-perturbative $\psi_{(1,3)}$
flow, using the results of \cite{Chim:1995kf,Lesage,Ahn:1998xm} that
have been obtained using TCSA and TBA techniques (see also
\cite{Fredenhagen:2002qn,Fredenhagen:2003xf}). For small label
$a_{2}$, the result is simply
\begin{equation}\label{noprule}
(1,a_{2})_{m-1} \quad \longrightarrow \quad 
(a_{2}-1,1)_{m-1} \ .
\end{equation}
Summarising our results so far, we predict the end-points of the RG flows 
to be
\begin{align}
\label{answ1}
(a_{1},1)_{m}\quad &\longrightarrow \quad (1,a_{1})_{m-1}\\
\label{answ2}
(1,a_{2})_{m} \quad & \longrightarrow \quad (a_{2}-1,1)_{m-1} \quad
\text{for}\ 1<a_{2} \ll m \ .
\end{align}

\noindent If we now want to consider the case of a general boundary
condition $(a_{1},a_{2})_{m}$, we face the problem that we 
do not know a priori how to identify the boundary field that couples to
$\tilde\mu$. One way to proceed would be to argue that the flow in question
must be the non-perturbative flow corresponding to the perturbative flow
from (II) to (III) that we have identified above; this will, as we shall see,
lead to the correct result. However, in the following we shall follow
a different route by studying the action of topological defect lines.

\subsection{Defect lines}

So far we have discussed how boundary conditions adjust themselves
under bulk deformations. We can also use similar methods to analyse what
happens to defect lines. Recall that defect lines interpolate between 
in general different conformal field
theories. In the following we shall only consider defects where the theories
on both sides are the same (namely some minimal model). 
More specifically, we shall consider topological defects 
that are characterised by the property that the corresponding defect
operator commutes with the action of both left- and right-moving Virasoro generators.
In particular, this implies that the defect only depends on the homotopy class
of the defect line \cite{Petkova:2000ip}. By moving the defect line to the
boundary ({\it i.e.}\ by `fusion'), such topological defects define then an action on 
the (conformal) boundary conditions of the theory.

For the charge conjugation theories we are considering here, the 
topological defects are labelled by the same labels as the Cardy
boundary states. We can therefore denote them as  
$\mathcal{D}(d_{1},d_{2})$, where 
${\bf d}=(d_1,d_2)$ has the same range and identification rules as the boundary
labels ${\bf a}=(a_1,a_2)$. The action of the defect $\mathcal{D}({\bf d})$ on
the boundary condition ${\bf a}$ is then described by the usual fusion rules
\begin{equation}
\mathcal{D}({\bf d}) \times {\bf a} = \bigoplus_{{\bf b}}
N_{{\bf d} {\bf a}}{}^{{\bf b}} \,\, {\bf b} \ .
\end{equation}
In addition to the identity defect $\mathcal{D} (1,1)$, there is a
special topological defect 
$\mathcal{D} (m-1,1)\cong \mathcal{D}(1,m)$ that generates a 
$\mathbb{Z}_{2}$ symmetry of the charge conjugation 
minimal model (such a defect is called a 'group-like defect'
in~\cite{Frohlich:2006ch} and a 'symmetry defect'
in~\cite{Brunner:2007ur}). Its action on a bulk field $\phi_{(r,s)}$
is given by
\begin{equation}\label{bulkfieldssymmetry}
\mathcal{D} (1,m) : \quad 
\phi_{(r,s)} \mapsto (-1)^{(m+1)r-ms+1} \phi_{(r,s)}\ ,
\end{equation}
which is indeed a symmetry of the bulk theory. Note that the field 
$\phi_{(1,3)}$ that we use to perturb the theory is invariant under
this $\mathbb{Z}_{2}$ action. The effect of the $\mathbb{Z}_{2}$
symmetry on a boundary condition is
\begin{equation}\label{symmetry}
(a_{1},a_{2}) \mapsto \mathcal{D} (1,m) \times (a_{1},a_{2}) =
(a_{1},m+1-a_{2}) \ .
\end{equation}
The concept of defect lines has proven very useful in the discussion
of boundary flows~\cite{Graham:2003nc,Bachas:2004sy} and combined bulk-boundary
flows~\cite{Brunner:2007ur}.
\smallskip

After these preparations we now want to study what happens to these
topological defects under the bulk deformation where we perturb the bulk theory 
on both sides of the defect by the $\phi_{(1,3)}$ bulk field. In principle,
this can also be analysed by the above methods since we can think of the defect
as a boundary condition in the doubled theory \cite{Oshikawa:1996dj}. 
Typically, the bulk deformation
will switch on defect fields --- the analogue of the boundary fields for boundary
conditions --- that will ensure that at the end of the flow the defect is again conformal. 
In general, however, a topological defect will not flow to a topological defect again,
and thus the identification of the end-point will be difficult. 

There is, however, one particularly simple class of defects for which
the analysis is essentially trivial. These are the defects
$\mathcal{D}(d_{1},1)$, on which no relevant defect operator can be
turned on by the $\phi_{(1,3)}$ perturbation. In fact, the spectrum of
a topological defect $\mathcal{D}({\bf d})$ is given by
\begin{equation}
\mathcal{H}_{\mathcal{D}({\bf d})} = 
\bigoplus_{\bf a,b,c} N_{\bf d c}{}^{\bf d} 
N_{\bf a b}{}^{\bf c}\, \mathcal{H}_{\bf a} \otimes \mathcal{H_{\bf b}} \ . 
\end{equation}
The defect spectrum can be decomposed under the action of the two 
Virasoro algebras. Thus each defect field $\psi_{\bf ab}$ 
carries two labels ${\bf a}=(a_{1},a_{2})$ and ${\bf b}=(b_{1},b_{2})$. For 
${\bf d}=(d_{1},1)$, we only find fields in the spectrum for which 
${\bf b}=(a_{1}+2n,a_{2})$ for some integer $n$. As the bulk field
$\phi_{(1,3)}$ can only turn on defect fields with $a_{1}=b_{1}=1$ and
$a_{2},b_{2}$ odd, we can only turn on the defect fields
$\psi_{(1,1+2p)(1,1+2p)}$ (for some integer $p$) 
on $\mathcal{D}(d_{1},1)$. Apart from the identity field, the field with lowest
conformal weight is $\psi_{(1,3)(1,3)}$ whose conformal weight is 
$h=2h_{(1,3)}=1+\frac{m-3}{m+1}$. This is greater than $1$ for $m>3$, and
thus the field is irrelevant. 

This argument shows that no relevant defect fields are 
turned on in our case. This should then imply that the
end-point of the flow is again a topological defect.\footnote{An independent argument 
for this defect to remain topological under the renormalisation group flow was
also very recently given in \cite{Kormos:2009sk}.} Under this
assumption it is then easy to determine to which topological defect
$\mathcal{D}(d_{1},1)_m$ flows in the $(m-1)^{\rm st}$ model. To
see how this goes let us consider a configuration with a defect line
$\mathcal{D}(d_{1},1)_{m}$ and the simplest boundary condition
$(1,1)_{m}$. Now we can either first act with the defect line on the
boundary and then do the bulk perturbation, or we can first analyse
the bulk perturbation when the defect and the boundary are far apart,
and then let the perturbed defect act on the perturbed boundary
condition.  If we first let the defect act on the boundary condition
in the $m^{\rm th}$ theory, we get the boundary condition
$(d_{1},1)_{m}$, which under the bulk perturbation flows to
$(1,d_{1})_{m-1}$ (see~\eqref{answ1}). If we do the bulk perturbation
first, we flow to a topological defect $\tilde{D}_{m-1}$ and the
boundary condition $(1,1)_{m-1}$. To have compatibility with the above
discussion, the action of the (topological) defect $\tilde{D}_{m-1}$
on $(1,1)_{m-1}$ must give $(1,d_{1})_{m-1}$; the only topological
defect that has this property is
$\tilde{\mathcal{D}}_{m-1}=\mathcal{D}(1,d_{1})_{m-1}$. Thus we
conclude that
\begin{equation}\label{defectflows}
\mathcal{D}(d_{1},1)_{m} \quad \longrightarrow \quad \mathcal{D}(1,d_{1})_{m-1} \ . 
\end{equation}
Note that this rule should be true for all values of $d_1$ since \eqref{answ1}
holds for all values of $a_1$. In particular, it 
also applies to the defect $\mathcal{D}(m-1,1)_{m}$ that
generates the $\mathbb{Z}_{2}$ symmetry. As the bulk perturbation by
$\phi_{(1,3)}$ is invariant under the $\mathbb{Z}_{2}$ symmetry, the
defect $\mathcal{D}(m-1,1)_{m}$ should flow to the $\mathbb{Z}_{2}$
generating defect $\mathcal{D}(m-2,1)_{m-1}\cong
\mathcal{D}(1,m-1)_{m-1}$ in the $(m-1)^{\rm st}$ theory, which is
precisely consistent with (\ref{defectflows}).

As another consistency check of this proposal we can consider the action
on bulk fields. From the analysis of~\cite{Zamolodchikov:1987ti} we know that
the $(r,s)$  bulk field in the $m^{\rm th}$ theory flows to a linear
combination $\phi '$ of bulk fields 
\begin{equation}
\phi_{(r,s)}^{(m)}\quad \longrightarrow \quad \phi ' = \sum_{s'-s\
\text{even}} c_{s'} \big[ \phi_{(s',r)}^{(m-1)}\big]\ ,
\end{equation} 
where $c_{s'}$ are some constants and we indicated by the square brackets that 
also descendants of the primary bulk fields can appear. On the other hand, we 
know the action of a topological defect $\mathcal{D} (d_{1},d_{2})_{m}$ on a bulk
field in the $m^{\rm th}$ theory (see {\it e.g.}\ \cite{Petkova:2000ip}),
\begin{equation}
\mathcal{D} (d_{1},d_{2})_{m}\,\, \phi_{(r,s)}^{(m)}  
= (-1)^{(d_{1}-d_{2}) (r-s)} 
\frac{ \sin \Bigl( \frac{\pi d_1 r}{m} \Bigr)}{ \sin \Bigl( \frac{\pi  r}{m} \Bigr)} \, 
\frac{ \sin \Bigl( \frac{\pi d_2 s}{m+1} \Bigr)}{ \sin \Bigl( \frac{\pi  s}{m+1} \Bigr)} \, 
\phi_{(r,s)}^{(m)} \ .
\end{equation}
If we first apply the defect $\mathcal{D} (d_{1},1)_{m}$ in the $m^{\rm th}$ theory, 
and then flow to the $(m-1)^{\rm st}$ theory we obtain 
\begin{equation}\label{defectonbulkI}
\mathcal{D} (d_{1},1)_{m}\, \, \phi_{(r,s)}^{(m)} =
(-1)^{(d_{1}-1) (r-s) }\, 
\frac{ \sin \Bigl( \frac{\pi d_1 r}{m} \Bigr)}{ \sin \Bigl( \frac{\pi  r}{m} \Bigr)} \,\,
\phi_{(r,s)}^{(m)} \quad
\longrightarrow \quad (-1)^{(d_{1}-1) (r-s) }
\frac{ \sin \Bigl( \frac{\pi d_1 r}{m} \Bigr)}{ \sin \Bigl( \frac{\pi  r}{m} \Bigr)} \,\,
\phi' \ .
\end{equation}
On the other hand, if we first let both the defect and the bulk field flow, we have to evaluate
\begin{equation}\label{defectonbulkII}
\mathcal{D} (1,d_{1})_{m-1}\,\, \phi ' =
\sum_{s'-s\ \text{even}} c_{s'} (-1)^{(d_{1}-1) (r-s')}\,
\frac{ \sin \Bigl( \frac{\pi d_1 r}{m} \Bigr)}{ \sin \Bigl( \frac{\pi  r}{m} \Bigr)} \,\,
\big[ \phi_{(s',r)}^{(m-1)}\big]  = 
(-1)^{(d_{1}-1)(r-s)} \frac{ \sin \Bigl( \frac{\pi d_1 r}{m} \Bigr)}{ \sin \Bigl( \frac{\pi  r}{m} \Bigr)} \,\,
\phi ' \ ,
\end{equation}
which precisely equals \eqref{defectonbulkI}, thus giving further support to the proposal
\eqref{defectflows}. 

\subsection{The final answer}

Putting everything together we can now determine what happens to an arbitrary
boundary condition $(a_{1},a_{2})_{m}$ ($1<a_{2}\ll m$)  under the bulk flow. To this end
we write it as the fusion of an appropriate topological defect on an
elementary boundary condition and then perform the flow under
the $\phi_{(1,3)}$ bulk perturbation,
\begin{align}
(a_{1},a_{2})_{m} & = \mathcal{D}(a_{1},1)_{m} \times (1,a_{2})_{m} \nonumber\\
 & \longrightarrow \mathcal{D}(1,a_{1})_{m-1} \times (a_{2}-1,1)_{m-1} =
(a_{2}-1,a_{1})_{m-1} \ . \label{answ2a}
\end{align}
Here we have assumed $1<a_2 \ll m$ since we have used~\eqref{answ2}. 
On the other hand $a_1$ is arbitrary because the defect argument 
did not depend on $a_{1}$ being small or not. If $a_2=1$ there is no 
relevant boundary field in the boundary spectrum, and we have instead
of (\ref{answ2a}) simply the flow~(\ref{answ1}). 

Note that we can also use the defect argument to check our identifications
for the perturbative fixed points (I), (II) and (III): if we apply the defects 
$\mathcal{D} (a_{1},1)_{m} \to \mathcal{D} (1,a_{1})_{m-1}$ to the
flow sequence for $a_1=1$ 
\begin{equation}
(1,a_{2})_{m} \stackrel{{\rm (I)}}{\longrightarrow} (a_{2},1)_{m} 
\stackrel{{\rm (II)}}{\longrightarrow} (1,a_{2})_{m-1} 
\stackrel{{\rm (III)}}{\longrightarrow} (a_{2},1)_{m-1} \quad \text{for}\ 1<a_{2}\ll m\ ,
\end{equation}
we find the sequence
\begin{equation}\label{sequencesmalla2}
(a_{1},a_{2})_{m} \stackrel{{\rm (I)}}{\longrightarrow} 
\bigoplus_{l} (b (l),1)_{m}
\stackrel{{\rm (II)}}{\longrightarrow}
\bigoplus_{l} (1,b (l))_{m-1}
\stackrel{{\rm (III)}}{\longrightarrow}
(a_{2},a_{1})_{m-1} \quad \text{for}\ 1<a_{2}\ll m\ ,
\end{equation}
where $b (l)=|a_{1}-a_{2}|-1+2l$ and $l=1,\dotsc ,\min (a_{1},a_{2},m-a_{1},m-a_{2})$.
This reproduces in particular our results from section~2 for small values
of $a_1$ and $a_2$. However, the current analysis is also true for
arbitrary values of $a_1$ since the rule for the defect flow is not restricted to small 
values  of $a_1$. 
\medskip

However, we still have the restriction  that $a_2$ should be small.
By using the identification rules 
$(a_{1},a_{2})_{m}\cong (m-a_{1},m+1-a_{2})_{m}$
we can also get a result for labels $a_{2}$ close to $m$. Firstly, we
notice that the flow~\eqref{answ1} translates into
\begin{equation}
(a_{1},m)_{m} \longrightarrow (1,m-a_{1})_{m-1} \ .
\end{equation}
For non-trivial label $a_{2}$, the sequence~\eqref{sequencesmalla2} of
the perturbative fixed-points (I), (II), (III) is mapped to
\begin{equation}\label{sequencelargea2}
(a_{1},a_{2})_{m} \stackrel{{\rm (I)}}{\longrightarrow} 
\bigoplus_{l} (b' (l),1)_{m}
\stackrel{{\rm (II)}}{\longrightarrow}
\bigoplus_{l} (1,b' (l))_{m-1}
\stackrel{{\rm (III)}}{\longrightarrow}
(a_{2}-2,a_{1})_{m-1} \quad \text{for}\ 1\ll a_{2} < m\ ,
\end{equation}
where now 
$b' (l)=|a_{1}-a_{2}+1|-1+2l$ and $l=1,\dotsc ,\min (a_{1},a_{2}-1,m-a_{1},m+1-a_{2})$. 
If we extrapolate this answer to small values of $a_2$, its form is different from that of 
the sequence~\eqref{sequencesmalla2}. The reason for this is that 
the notion of which fixed points are perturbative and which are non-perturbative 
changes as we extrapolate $a_{2}$ from small to large values. In particular, instead
of the perturbative pure boundary fixed point (\ref{RRS}) the first flow
(to fixed point (I)) is now of the same form as 
the non-perturbative boundary flow (that generalises (\ref{noprule}))
\begin{equation}\label{Ip}
(a_1,a_2) \longrightarrow 
\bigoplus_{l=1}^{\min (a_{1},a_{2}-1)} (a_{1}+a_{2}-2l,1) \ ,
\end{equation}
see~\cite{Chim:1995kf,Lesage,Ahn:1998xm,Recknagel:2000ri}.
Similarly, the third flow (from (II) to (III)) has now the same structure
as the non-perturbative analogue of (\ref{IItoIII}), see
\cite{Graham:2001pp,Fredenhagen:2001kw,Fredenhagen:2002qn,Fredenhagen:2003xf}.

Remarkably, however, the
form of the actual fixed-point (IV) does not change for large values
of $a_{2}$, and we find
\begin{equation}\label{answ2b}
(a_{1},a_{2})_{m} \longrightarrow (a_{2}-1,a_{1})_{m-1} \qquad
\text{for}\ 1\ll a_{2} < m -1 \ ,
\end{equation}
which is the same formula as~\eqref{answ2a}.  This suggests that the
result for the end point --- the fixed point (IV) ---  of the actual flow can 
be interpolated to intermediate values of  $a_{2}$; a rather non-trivial
test for this conjecture will be presented in section~4.4.
Thus we are led  to the final answer for the bulk-boundary flow 
\begin{equation}\label{answf}
\boxed{(a_1,a_2)_m \quad \longrightarrow \quad \left\{
\begin{array}{cl}
(1,a_1)_{m-1} & \qquad \hbox{if $a_2=1$} \\[4pt]
(a_{2}-1,a_{1})_{m-1}& \qquad
\hbox{if $1<a_2<m$} \\[4pt] 
(1,m-a_{1})_{m-1}& \qquad \hbox{if $a_{2}=m$}
\end{array}
\right.}
\end{equation}
under the $\phi_{(1,3)}$ bulk perturbation. 
We should mention that (\ref{answf}) reproduces in particular the conjectured flow
of \cite{Lesage,Ahn:1998xm},
\begin{equation}
(1,a)_{m} \longrightarrow (a-1,1)_{m-1} 
\end{equation}
for $1<a<m$.
\smallskip

Finally, let us comment on the behaviour of the flows under the 
$\mathbb{Z}_{2}$ symmetry of the minimal models (see~\eqref{symmetry}). 
The bulk field $\phi_{(1,3)}$ is invariant under this symmetry 
(see~\eqref{bulkfieldssymmetry}), and thus we
expect the flows and fixed-points to respect the 
symmetry.\footnote{One can explicitly check that the coupled RG equations remain
invariant under the symmetry. There is a slight subtlety here,
namely that the normalisation of the boundary fields in
\cite{Runkel:1998pm} --- this has an impact on the OPE coefficients
that enter this calculation, see appendix~A --- depend on $m$. Once
this has been taken into account, we find the same structures for
the case of small $a_1$ and $a_2$ and for the $\mathbb{Z}_{2}$-image of
this case.} In fact, it is not difficult to see that under the map
$(a_1,a_2)_m\mapsto (a_1,m+1-a_2)_m$ the two sequences 
\eqref{sequencesmalla2} and~\eqref{sequencelargea2} 
are mapped into one another, and that the 
results~\eqref{answ2a} and~\eqref{answ2b} on the actual
fixed-point (IV) are left invariant.

\subsection{Consistency constraints}

Finally, there are a number of consistency constraints that we can
check. They are all a variant of the following
observation.\footnote{We thank Patrick Dorey for drawing our attention
to this point.}  The flows from our initial boundary condition to the
fixed-point (IV) come in a one-parameter family labelled by a
parameter $\chi$ that measures the initial strength of the boundary
perturbation by $\psi_{(1,3)}$ relative to the bulk perturbation. (More
precisely $\chi$ is proportional to the initial value of the ratio 
$\mu /\sqrt{|\lambda |}$ when the perturbation is turned on,
see~\eqref{meaningofchi} in appendix~C.)

For large negative values of $\chi$, the flow first follows closely
the pure boundary flow to (I), then the bulk flow down to (II), and
finally the pure boundary flow from (II) to (IV) (see
figure~\ref{fig:flowgraph}). In the limit $\chi \to -\infty$, the flow
completely decomposes into three separate flows. If our analysis is
correct, the separate flows should involve the same fixed-points that
we identified in our analysis. The fixed point (I) that is reached by
the pure boundary perturbation is described by a superposition of
boundary conditions $(b,1)$. The subsequent bulk flow is thus given by
(\ref{a2=1}). Combining these two flows, {\it i.e.} (\ref{RRS}) and
(\ref{a2=1}) then leads indeed to (\ref{II}) with the same
configuration at (II).

Similarly we can consider the limit $\chi \to +\infty$. Then we first
follow a pure boundary flow that is the non-perturbative counterpart
of the perturbative flow to (I), and we reach a fixed point (I') that
corresponds to the superposition (\ref{Ip}). For the subsequent bulk
flow we can again use~\eqref{a2=1} and we arrive at a superposition (II') of
boundary conditions of the form $(1,b_{i})$. From there we can follow
a pure boundary flow. Although we do not know a priori which boundary
$(1,3)$-fields are turned on, we know that at least there is a
perturbative boundary flow (see \eqref{kevinsflows})
involving in general also boundary condition changing fields that
leads to the fixed point (IV) \cite{Graham:2001pp}, see figure~2.   
To ensure that this is really the flow that takes place, one can again
consider the case of $a_{1}=1$ where there is only one perturbative boundary
(1,3)-flow from (II'), and so it is easily identified. To get the flow
for the general case, one then again uses the fusion of a defect~\eqref{defectflows}. 

\begin{figure}[htb]\begin{center}
\includegraphics{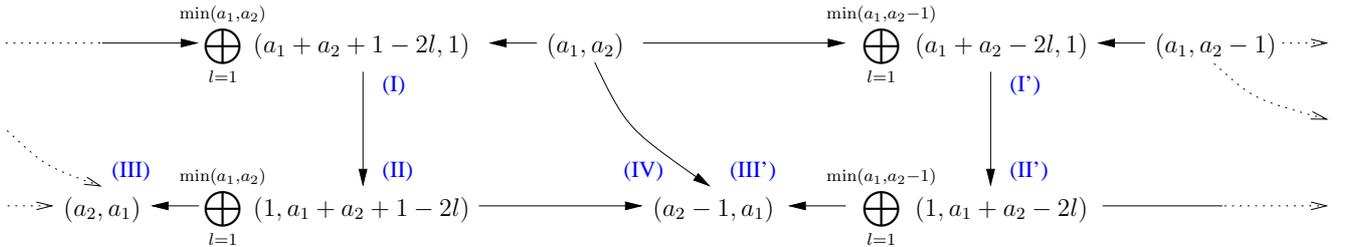}
\end{center}
\caption{\label{fig:flowchain}A segment of the chain of flows for
small values of $a_{1},a_{2}$. In the upper line we have boundary
conditions of the $m^{\rm th}$ and in the lower line of the 
$(m-1)^{\rm st}$ minimal model.}
\end{figure}

This last boundary flow in the $c_{m-1}$ theory is actually the
perturbative flow from (II') to (III') for another initial boundary
condition ${\bf a'}=(a_{1},a_{2}-1)$. We can therefore combine the 
flow graphs for all boundary conditions $(a_{1},\cdot)$ into a chain
of flows; figure~\ref{fig:flowchain} shows a generic segment of the
chain. 

\begin{figure}[htb]\begin{center}
\includegraphics{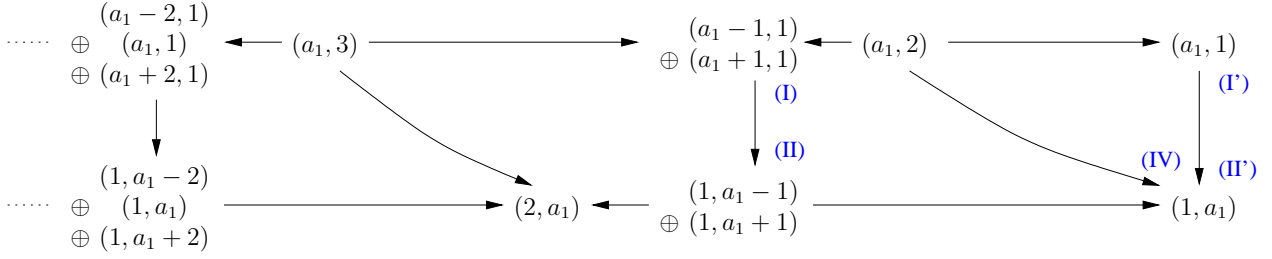}
\end{center}
\caption{\label{fig:flowchain-end}The right end of the chain of
flows. Starting from the boundary condition~$(a_{1},2)_{m}$, in one
direction the flow decomposes into only two separate flows ($\to
(a_{1},1)_{m}\to (1,a_{1})_{m-1}$) instead of three separate flows in
the generic case.}
\end{figure}

At the end of the chain the situation degenerates: the boundary
condition $(a_{1},2)_{m}$ flows to $(1,a_{1})$, so the fixed-point
(IV) coincides with the fixed-point (II') that is reached from the
pure bulk flow from (I'), see figure~\ref{fig:flowchain-end}.

The chains that we get for different values of $a_{1}$ look all
similar. Indeed, the chain for general $a_1$ can be obtained starting from 
the chain with $a_{1}=1$ by fusion with a topological defect. The
topological defect to consider is the one that corresponds to
$\mathcal{D} (a_{1},1)_{m}$ in the $m^{\rm th}$ model. Along the upper 
sequence ({\it i.e.}\ in the $m^{\rm th}$ theory) this topological defect 
maps indeed the upper part of the $a_1=1$ chain to the upper
part of the chain for $a_1$. As we have
argued before, the defect $\mathcal{D} (a_{1},1)_{m}$ flows 
under the bulk flow to $\mathcal{D} (1,a_{1})_{m-1}$, and it is 
therefore this topological defect that acts on the lower sequence. This then
produces the lower sequence of the diagram for $a_1$. 

\subsection{A perturbative analysis in the middle of the chain of flows}

\begin{figure}[htb]\begin{center}
\includegraphics{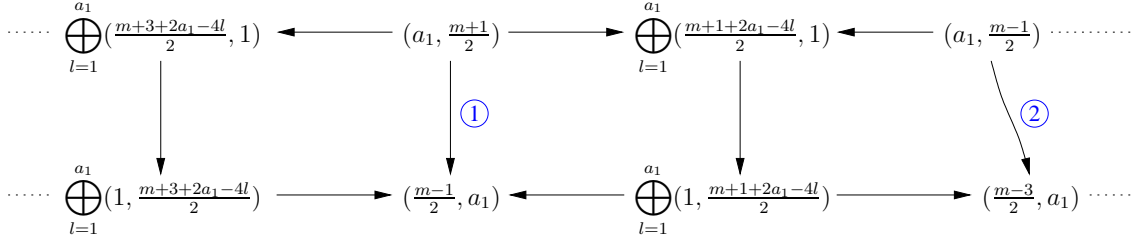}
\end{center}
\caption{\label{fig:flowchain-mid}The middle of the chain of
flows for $m$ odd. The fixed point starting from the exact middle
$(a_{1},\frac{m+1}{2})_{m}$ is reached for $\lambda =\lambda_{*}$,
$\mu =0$ (flow~1 in the figure). 
When we go slightly away from the middle
(flow~2), the flow can still be treated perturbatively.}
\end{figure}

When we extrapolate our chain from the right to larger values of
$a_{2}$, the perturbative boundary flows become longer, and the
non-perturbative flows become shorter. In the middle of the chain,
they are of the same length, more precisely, for the boundary
condition $(a_{1},\frac{m+1}{2})_{m}$ ($m$ odd), the two boundary
flows triggered by $\psi_{(1,3)}$ are mapped to each other by the
$\mathbb{Z}_{2}$ symmetry and are thus equally long (see
figure~\ref{fig:flowchain-mid}). The diagram suggests that for this
boundary condition the pure bulk perturbation does not switch on the
boundary field $\psi_{(1,3)}$. Indeed one finds that the bulk boundary
coefficient $B_{(1,3)}^{\;(1,3)}=0$, so the fixed-point is in this case
given by $\lambda =\lambda_{*}$ and $\mu =0$. We can then compute the
perturbed $g$-function, which to leading order only gets a
contribution from the bulk one-point function. We find 
\begin{equation}
\log \frac{g_{\lambda_{*},0}^{(m)} (a_{1},\frac{m+1}{2})}{g^{(m)}
(a_{1},\frac{m+1}{2})} = \frac{1}{4m}\frac{2-3\xi}{\xi} + \mathcal{O}
(\xi,m^{-2}) \ .
\end{equation}
With our prescription for the cutoff $\xi$ (see~\eqref{etapres}), we
obtain the expected result
\begin{equation}
\log \frac{g^{(m-1)} (\frac{m-1}{2},a_{1})}{g^{(m)}
(a_{1},\frac{m+1}{2})} = -\frac{1}{2m} + \mathcal{O} (m^{-2})\ .
\end{equation}
This check gives support to the idea that our results can 
be extrapolated to any value of $a_{2}$. We can go even further and
look at boundary conditions close to
$(a_{1},\frac{m+1}{2})_{m}$. Figure~\ref{fig:flowchain-mid} suggests
that here the bulk boundary flow to the true fixed point (IV) is perturbative
whereas the pure boundary flows are nonperturbative.
To check this explicitly, let us consider the RG equations for the boundary coupling 
$\mu$ of the $\psi_{(1,3)}$ field for a boundary condition of the form
${\bf a}=(a_{1},\tfrac{m+1}{2}-a_{2})$,
\begin{align}
\dot{\mu}&= (1-h)\mu  + \frac{1}{2}B\lambda + E\lambda \mu + \dotsb \nonumber\\
 &= \frac{2}{m}\mu -\pi a_{2}\lambda +4\pi \lambda \mu + \dotsb \ ,
\label{RGeqmiddle}
\end{align}
where $a_1$ and $a_2$ are now small. Here we have used the expressions 
for the coupling constants from appendix~A.2. Note that we neglected the term 
proportional to $\mu^{2}$, because its coefficient $\tri{D}{(1,3)}{(1,3)}{(1,3)}$ is 
suppressed by $m^{-3}$. For the calculation of $E$ we need to determine
the correlator of the bulk field and two boundary fields to leading order in $1/m$,
which now takes the form
\begin{equation}\label{correlator1}
\frac{1}{\cor{\id} \tri{D}{(1,3)}{(1,3)}{(1,1)}}\, 
\cor{\phi_{(1,3)}(z,\bar{z})%
\psi_{(1,3)}(0)\psi_{(1,3)}(\infty)}=
\frac{ - \frac{3}{2} (z-\bar{z})^4 + \frac{1}{2} (z+\bar{z})^4}
{|z|^4(z-\bar{z})^2}+\mathcal{O}(m^{-1})\ .
\end{equation}
In particular, this correlator has the correct asymptotics in the various
channels:  for the  $(1,1)$ and $(1,5)$ channels the relevant coefficients are
now
\begin{equation}
B_{(1,3)}^{(1,1)} D_{(1,1)(1,3)}^{\phantom{(1,5)}(1,3)} = 
-1+\mathcal{O}(\tfrac{1}{m^2}) \ , \qquad 
B_{(1,3)}^{(1,5)} D_{(1,5)(1,3)}^{\phantom{(1,5)}(1,3)} = \tfrac{4}{3} 
+\mathcal{O}(\tfrac{1}{m}) \ ,
\end{equation}
whereas the coefficient in the $(1,3)$ channel still vanishes to leading order 
\begin{equation}
B_{(1,3)}^{(1,3)} D_{(1,3)(1,3)}^{\phantom{(1,5)}(1,3)} = {\cal O}(m^{-3}) \ .
\end{equation}
After integration as in (\ref{E_raw}) this leads to $E=4\pi$. This is the expected
value since at the fixed point
$\lambda=\lambda_*=-\frac{1}{\pi m}$ the conformal dimension of the field
coupling to $\mu$ should be $h_{(3,1)} = 1 + \tfrac{2}{m}+\cdots$. 

\begin{figure}[htb]\begin{center}
\resizebox{13cm}{!}{\includegraphics{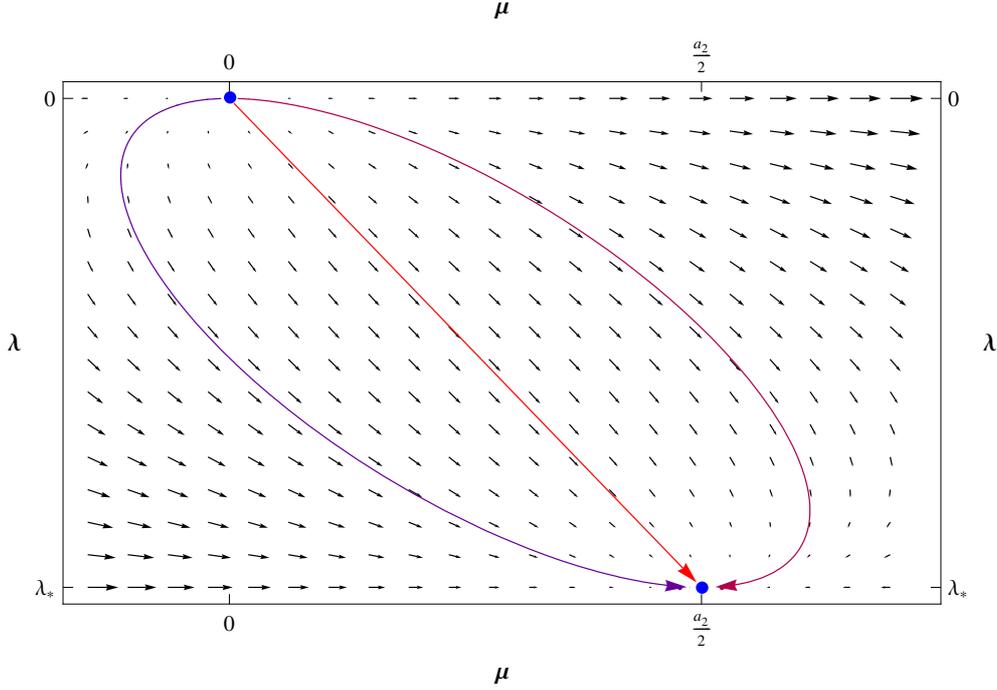}}
\end{center}
\caption{\label{fig:flowgraph2}The flow diagram
for~\eqref{RGeqmiddle} for $a_{2}=1$ and $m=101$. The vectors
$(\dot{\mu},\dot{\lambda})$ have been magnified by a factor $2.5$. A
generic flow (described by the solution~\eqref{RGsolutionmiddle}
of~\eqref{RGeqmiddle}) reaches the perturbative fixed-point.}
\end{figure}

For $\lambda =\lambda_{*}$ the RG equation~\eqref{RGeqmiddle} has a
unique fixed point at $\mu_{*}=\frac{a_{2}}{2}$. This solution
describes a {\em perturbative} fixed point. In fact the order one
value of $\mu_*$ is only an artefact of our normalisation since
$D_{(1,3)(1,3)}^{\phantom{(1,3)}(1,1)}$ is of order $m^{-2}$, and thus
in the `natural normalisation' we should rescale the field $\psi$ by a
factor of $m$, which would replace $\mu\mapsto \frac{\mu}{m}$. Note
that after this rescaling the boundary coupling constant
$\tri{D}{(1,3)}{(1,3)}{(1,3)}$ is only of order $m^{-2}$, and hence can
still be ignored in the RG equation.  It is also clear from the flow
diagram (see figure~\ref{fig:flowgraph2}) that this fixed point is
actually reached by a generic flow.

Using the same techniques as above in section~3, we can now determine the  
change in the $g$-factor,
\begin{align}
\log g^{(m)}_{\lambda_{*},\mu_{*}} ({\bf a}) 
&= \log g^{(m)}({\bf a}) -\frac{16\pi^{2}}{m^{3}}\mu_{*}^{2} -
\frac{2\pi^{3}a_{2}^{2}}{m^{2}}\lambda_{*}\bigg(1-\frac{8}{a_{2}}\mu_{*}
+\frac{16}{a_{2}^{2}}\mu_{*}^{2} \bigg) + \dotsb \nonumber\\
& = \log g^{(m)}({\bf a})
-\frac{2\pi^{2}a_{2}^{2}}{m^{3}} + \dotsb \ ,
\end{align}
where the dots denote contributions that are either of higher order in
$1/m$ or do not depend on the boundary labels. To remove the terms
that are independent of the boundary labels, we again look at relative
$g$-functions with respect to some reference boundary condition, which
we choose here to be $(1,\frac{m+1}{2})_{m}$ (notice that we have to use a
boundary condition which is in the regime where the perturbative
analysis applies). The relative $g$-function in the $m^{\rm th}$ model is
\begin{equation}
\frac{g^{(m}({\bf  a)}}{g^{(m)}(1,\frac{m+1}{2})} = \frac{\sin \frac{\pi a_{1}}{m}}{\sin \frac{\pi}{m}}
\bigg(1 - \frac{\pi^{2}a_{2}^{2}}{2m^{2}} +
\frac{\pi^{2}a_{2}^{2}}{m^{3}} + \mathcal{O}(m^{-4})  \bigg) \ .
\end{equation}
After the perturbation by $\phi_{(1,3)}$ the relative $g$-function becomes
\begin{align}
\log \frac{g^{(m)}_{\lambda_{*},\mu_{*}} ({\bf
a})}{g^{(m)}_{\lambda_{*},0}(1,\frac{m+1}{2})} &= \log
\frac{\sin \frac{\pi a_{1}}{m}}{\sin \frac{\pi}{m}} - \frac{\pi^{2}a_{2}^{2}}{2m^{2}} -
\frac{\pi^{2}a_{2}^{2}}{m^{3}} + \mathcal{O}(m^{-4}) \\
&= \log
\frac{g^{(m-1)}(\frac{m-1}{2}-a_{2},a_{1})}{g^{(m-1)}(\frac{m-1}{2},1)} 
+ \mathcal{O}(m^{-4}) \ .
\end{align}
This perturbative calculation thus predicts that the end-point of the actual
flow is
\begin{equation}\label{middle}
\left(a_1,\frac{m+1}{2}-a_2\right)_m \quad \stackrel{{\rm (IV)}}{\longrightarrow} 
\quad  \left(\frac{m-1}{2}-a_2,a_1\right)_{m-1}  \ .
\end{equation}
It is remarkable that in this case the actual end-point of the flow 
({\it i.e.} the fixed point (IV)) can be directly computed perturbatively. 
The result (\ref{middle}) is in beautiful agreement with our general answer
(\ref{answf}), and thus gives strong support for the claim that (\ref{answf})
is also true for intermediate values of $a_2$.

\section{Comparison with numerical results}
\setcounter{equation}{0}

Finally let us analyse how our perturbative calculation compares with
numerical calculations that have been performed before. In particular,
in \cite{Pearce:2000dv, Pearce:2003km} the flow of the tricritical
Ising ($m=4$) to the Ising model ($m=3$) was considered in the
presence of boundaries. They considered a cylinder diagram with two
boundaries for which they imposed the boundary conditions $(r,1)$ and
$(1,s)$, respectively.  According to the fusion rules, the relative
open string spectrum between the two boundary conditions transforms
then in the $(r,s)$ representation, and they found that under the
$\phi_{(1,3)}$ flow, the open string character $\chi_{(r,s)}$ flowed
as\footnote{Notice that there is a typographical error in Table~1
of~\cite{Pearce:2003km}. As explained in the accompanying text
of~\cite{Pearce:2003km}, the flow of the characters $\chi_{(1,2)}$
and $\chi_{(1,3)}$ in the $m=4$ model is rather as indicated
in~\eqref{characterflow}. We thank Leung Chim for pointing this out to
us.}
\begin{equation}\label{characterflow}
\begin{array}{ccl}
m=4&&m=3\\[4pt]
\chi_{(1,1)}&\,\longrightarrow\,&\chi_{(1,1)} \\
\chi_{(2,1)}&\,\longrightarrow\,&\chi_{(1,2)} \\
\chi_{(3,1)}&\,\longrightarrow\,&\chi_{(1,3)}\\
\chi_{(1,2)}&\,\longrightarrow\,&\chi_{(1,1)}\\
\chi_{(1,3)}&\,\longrightarrow\,&\chi_{(2,1)}\\
\chi_{(2,2)}&\,\longrightarrow\,&\chi_{(1,2)}\ .
\end{array}
\end{equation}
In particular, the first flow implies that the $(1,1)$ boundary
condition of the $m=4$ theory flows to the $(1,1)$ boundary condition
of the $m=3$ theory. In order to compare with our calculations (where
we consider a single boundary condition) we can take either $r=1$ or
$s=1$ since then one of the two boundary conditions does not flow. In
this way we deduce from (\ref{characterflow}) that the boundary
conditions flow as
\begin{equation}\label{boundaryprediction}
\begin{array}{c@{\quad \longrightarrow\quad}l}
(1,1)_4 &(1,1)_3\\
(2,1)_4 &(1,2)_3 \\
(3,1)_4 &(1,3)_3 \\
(1,2)_4 &(1,1)_3 \\
(1,3)_4&(2,1)_3\cong (1,3)_3\ .
\end{array}
\end{equation}
Unfortunately, we cannot determine the flow of the $(2,2$) boundary
condition in this manner, since none of the cylinder diagrams
considered in \cite{Pearce:2000dv, Pearce:2003km} involves this
boundary condition. However, one may suspect that the behaviour of the
cylinder diagram between $(1,1)$ and $(2,2)$ should be the same as that
between $(1,2)$ and $(2,1)$. Following this line of argument would lead to 
the prediction $(2,2)_4\rightarrow (1,2)_3$, which is indeed in agreement with our 
general rule~\eqref{answf}.

The flows (\ref{boundaryprediction}) agree perfectly with our general 
rule~\eqref{answf}. A schematic view of the flows in the space of couplings 
$\lambda ,\mu$ is as follows:
\begin{center}
\includegraphics{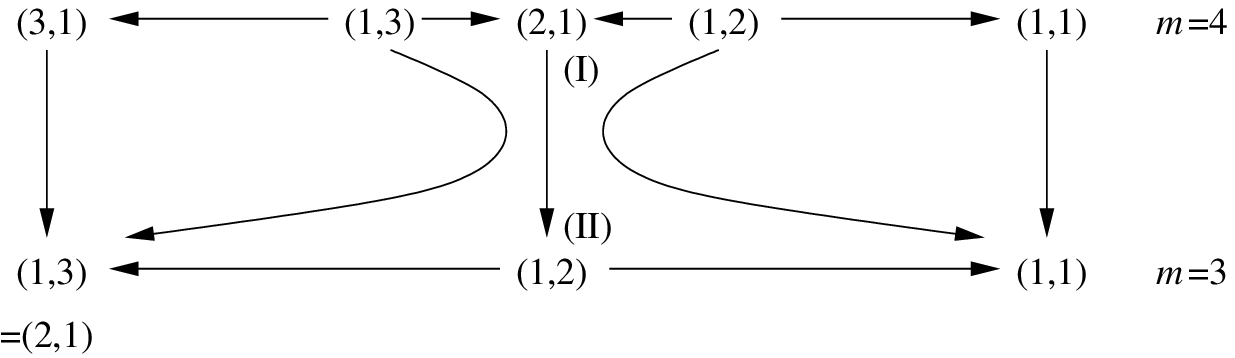}
\end{center}
The three vertical flows (no boundary fields are switched on) are the
first three flows in~\eqref{boundaryprediction}. The horizontal flows
are the well known boundary flows in the $m=4$ and $m=3$ model,
respectively, which are driven by the $\psi_{(1,3)}$ boundary
field. The short horizontal arrows in the $m=4$ model indicate that
these flows become the {\em perturbative} $\psi_{(1,3)}$ flows when
$m$ is sent to large values.\footnote{The flow $(1,2)_{4}\to
(2,1)_{4}$ becomes $(1,2)_{m}\to (2,1)_{m}$ for large $m$, while the
flow $(1,3)_{4}\to (2,1)_{4}$ becomes $(1,m-1)_{m}\to (m-2,1)_{m}$.}
Finally, the flows indicated by the curved arrows agree precisely with
the last two flows of \eqref{boundaryprediction}.

This final check of our general prediction~\eqref{answf} at this low value
of $m$ makes us confident that our results hold for the whole (A-type) series
of unitary minimal models.

\section*{Acknowledgements}

We are indebted to Patrick Dorey, Anatoly Konechny and Ingo Runkel for detailed 
explanations and discussions. We also thank Albion Lawrence 
and Volker Schomerus for initial collaboration on this project, and Andreas
Recknagel for useful comments.
The work of MRG and CSC is supported by the 
Swiss National Science Foundation. MRG thanks the GGI in Florence, the 
Chinese Academy of
Sciences in Beijing, and the IPMU in Tokyo for hospitality during the final stages 
of this work.

\begin{appendix}
\setcounter{section}{0}
\renewcommand{\thesection}{\Alph{section}}
\setcounter{equation}{0}
\renewcommand{\theequation}{\Alph{section}.\arabic{equation}}

\section*{Appendix }

\section{OPE coefficients}

For the charge conjugation minimal model (the A-series), the various 
OPE coefficients have been determined in  \cite{Runkel:1998pm} in terms of the 
$F$ matrices.  For large $m$ the bulk coupling constant $\tri{C}{(1,3)}{(1,3)}{(1,3)}$
has the following large $m$ behaviour
\begin{equation}
\tri{C}{(1,3)}{(1,3)}{(1,3)}=4-\frac{12}{m}-
\frac{8}{3}(3+\pi^2)\frac{1}{m^2}+\mathcal{O}(m^{-3})\ . 
\end{equation}
We also need the large $m$ expansions of the boundary and bulk-boundary OPE 
coefficients. 

\subsection{The expansion for small $a_1,a_2$}

For a boundary condition $a=(a_1,a_2)$ for which $a_1,a_2\ll m$, we have the following asymptotic expansions of the OPE coefficients:
\begin{eqnarray}
\tri{D}{(1,1)}{(1,3)}{(1,3)}&=& 1 \ ,
 \\[4pt]
\tri{D}{(1,3)}{(1,3)}{(1,1)}&=&\left\{
\begin{array}{ll}
\frac{2(a_2-1)}{(a_1-a_2)^2(a_2+1)}+\mathcal{O}(\frac{1}{m})&(a_2>a_1)\\[4pt]
\frac{2m^2}{(a_2^2-1)}-\frac{6m}{a_2^2-1}+\frac{4}{a_2^2-1}
+\mathcal{O}(\frac{1}{m})&(a_2=a_1)\\[4pt]
\frac{2(a_2+1)}{(a_1-a_2)^2(a_2-1)}+\mathcal{O}(\frac{1}{m})&(a_2<a_1) \ ,
\end{array}\right. \\
\tri{D}{(1,3)}{(1,3)}{(1,3)}&=&\left\{
\begin{array}{ll}
- \frac{4}{(a_1-a_2)(a_2+1)}+\mathcal{O}(\frac{1}{m})&(a_2>a_1)\\[4pt]
-\frac{4m}{(a_2^2-1)}+\mathcal{O}(1)&(a_2=a_1)\\[4pt]
- \frac{4}{(a_1-a_2)(a_2-1)}+\mathcal{O}(\frac{1}{m})&(a_2<a_1) \ ,
\end{array}\right. \\[8pt]
\tri{D}{(1,5)}{(1,3)}{(1,3)}&=&\left\{
\begin{array}{ll}
-\frac{24(a_2-2)}{(a_2+1)(a_1-a_2)^2((a_1-a_2)^2-1)}%
+\mathcal{O}(\tfrac{1}{m})                                      &(a_2\geq a_1+2)\\[4pt] 
\frac{12m}{a_1+2}+\frac{2(15a_1-34)}{a_1+2}%
+\mathcal{O}(\tfrac{1}{m})                                          &(a_2=a_1+1)\\[4pt]
\frac{24m^2}{a_1^2-1}-\frac{172m}{a_1^1-1}%
+\frac{4(6a_1^2+77+12\pi^2)}{a_1^2-1}+\mathcal{O}(\tfrac{1}{m})     &(a_2=a_1)\\[4pt]
-\frac{12m}{a_1-2}+\frac{2(15a_1+34)}{a_1-2}%
+\mathcal{O}(\tfrac{1}{m})                                          &(a_2=a_1-1)\\[4pt]
-\frac{24(a_2+2)}{(a_2-1)(a_1-a_2)^2((a_1-a_2)^2-1)}%
+\mathcal{O}(\tfrac{1}{m})					&(a_2\leq a_1-2) \ ,
\end{array}\right. \\[8pt]
\tri{D}{(1,5)}{(1,5)}{(1,1)}&=&\left\{
\begin{array}{ll}
\frac{864(a_2-1)(a_2-2)}{(a_2+1)(a_2+2)%
(a_1-a_2)^4((a_1-a_2)^2-1)^2}+\mathcal{O}(\tfrac{1}{m})		&(a_2\geq a_1+2)\\[4pt]
\frac{216 a_1\,m^2}{(a_1+3)(a_1+2)(a_1-1)}+\mathcal{O}(m)           &(a_2=a_1+1)\\[4pt]
\frac{864 m^4}{(a_1+2)(a_1+1)(a_1-1)(a_1-2)}+\mathcal{O}(m^3)	&(a_2=a_1)\\[4pt]
\frac{216 a_1\,m^2}{(a_1+1)(a_1-2)(a_1-3)}+\mathcal{O}(m)	    &(a_2=a_1-1)\\[4pt]
\frac{864(a_2+1)(a_2+2)}{(a_2-1)(a_2-2)%
(a_1-a_2)^4((a_1-a_2)^2-1)^2}+\mathcal{O}(\tfrac{1}{m})		&(a_2\leq a_1-2) \ ,
\end{array}\right.
\\[8pt]
\tri{D}{(1,5)}{(1,5)}{(1,3)}&=&\left\{
\begin{array}{ll}
-\frac{5184(a_2-2)}{((a_1-a_2)^2-1)^2(a_1-a_2)^3(a_2+1)(a_2+2)}
+\mathcal{O}(\tfrac{1}{m})															&(a_2\geq a_1+2)\\[4pt]
\frac{1296m^2}{(a_1+3)(a_1+2)(a_1-1)}+\mathcal{O}(m)		&(a_2=a_1+1)\\[4pt]
-\frac{5184m^3}{(a_1+2)(a_1+1)(a_1-1)(a_1-2)}
+\mathcal{O}(m^2)																				&(a_2=a_1)\\[4pt]
-\frac{1296m^2}{(a_1-3)(a_1-2)(a_1+1)}+\mathcal{O}(m)		&(a_2=a_1-1)\\[4pt]
-\frac{5184(a_2+2)}{((a_1-a_2)^2-1)^2(a_1-a_2)^3(a_2-1)(a_2-2)}
+\mathcal{O}(\tfrac{1}{m})															&(a_2\leq a_1-2) \ ,
\end{array}\right.
\\[8pt]
B_{(1,3)}^{\;(1,1)}&=&3-\frac{4\pi^2}{m^2}a_2^2+\mathcal{O}(\frac{1}{m^3})\ ,\\[8pt]
B_{(1,3)}^{\;(1,3)}&=&\left\{\begin{array}{ll}
\frac{4\pi}{m}(a_1-a_2)(a_2+1)+\mathcal{O}(\frac{1}{m^2})&(a_2>a_1)\\[4pt]
\frac{4\pi}{m^2}(a_2^2-1)+\mathcal{O}(\frac{1}{m^3})&(a_2=a_1)\\[4pt]
\frac{4\pi}{m}(a_1-a_2)(a_2-1)+\mathcal{O}(\frac{1}{m^2})&(a_2<a_1) \ ,
\end{array}\right. 
\\[8pt]
B_{(1,3)}^{\;(1,5)}&=&\left\{%
\begin{array}{ll}
-\frac{\pi^2}{18m^2}(a_2+1)(a_2+2)(a_1-a_2)^2((a_1-a_2)^2-1)%
+\mathcal{O}(\tfrac{1}{m^3})                                        &(a_2\geq a_1+2)\\[4pt]
\frac{\pi^2}{9m^3}(a_1-1)(a_1+2)(a_1+3)+\mathcal{O}(\tfrac{1}{m^4}) &(a_2=a_1+1)\\[4pt]
\frac{\pi^2}{18m^4}(a_1^2-4)(a_1^2-1)+\mathcal{O}(\tfrac{1}{m^5})   &(a_2=a_1)\\[4pt]
-\frac{\pi^2}{9m^3}(a_1+1)(a_1-2)(a_1-3)+\mathcal{O}(\tfrac{1}{m^4})&(a_2=a_1-1)\\[4pt]
-\frac{\pi^2}{18m^2}(a_2-1)(a_2-2)(a_1-a_2)^2((a_1-a_2)^2-1)%
+\mathcal{O}(\tfrac{1}{m^3})                                        &(a_2\leq a_1-2) \ .
\end{array}\right.
\end{eqnarray}

\subsection{Asymptotic expansion for $a_1$ small and $a_2$ close to 
$\frac{m+1}{2}$}

For boundary conditions of the form $(a_1,\tfrac{m+1}{2}-a_2)_m$ where
$m$ is odd and $a_1,a_2\ll m$ we have on the other hand
\begin{eqnarray}
\tri{D}{(1,3)}{(1,3)}{(1,1)}&=&\frac{8}{m^2}+\mathcal{O}(\tfrac{1}{m^3})\ ,\\[4pt]
\tri{D}{(1,3)}{(1,3)}{(1,3)}&=&\frac{8\pi^2a_2}{m^3}+\mathcal{O}(\tfrac{1}{m^4})\ ,
\\[4pt]
\tri{D}{(1,1)}{(1,3)}{(1,3)}&=&1\ ,
\\[4pt]
\tri{D}{(1,5)}{(1,3)}{(1,3)}&=&
-\frac{384}{m^4}+\mathcal{O}(\tfrac{\log m}{m^5})\ ,
\\[4pt]
\tri{D}{(1,5)}{(1,5)}{(1,1)}&=&\frac{9\cdot 2^{13}}{m^8}+
\mathcal{O}(\tfrac{\log m}{m^9})\ ,
\\[4pt]
\tri{D}{(1,5)}{(1,5)}{(1,3)}&=&2^{13}\cdot3^4\,\pi^2\,a_2\,\frac{1}{m^9}+
\mathcal{O}(\tfrac{\log m}{m^{10}})\,,\\[4pt]
B_{(1,3)}^{\;(1,1)}&=&-1+\frac{4\pi^2a_2^2}{m^2}+\mathcal{O}(\tfrac{1}{m^3})\ ,\\[4pt]
B_{(1,3)}^{\;(1,3)}&=&-2\pi a_2+\mathcal{O}(\tfrac{1}{m})\ ,\\[4pt]
B_{(1,3)}^{\;(1,5)}&=&-\frac{m^4}{288}+\mathcal{O}(\log(m)\,m^3)\ .
\end{eqnarray}

\section{Correlation functions and integrals}
\setcounter{equation}{0}
\subsection{Disc and upper half plane}

The upper half plane is defined by 
$\uhp=\{z\in\mathbb{C}\;|\;\im z\geq0\}$, while
the disc is defined by 
$\disc=\{w\in\mathbb{C}\;|\;|w|\leq 1\}$. On the disc
we often use the polar coordinates $w=re^{i\theta}$, where
$0\leq r\leq 1$ and $-\pi<\theta\leq\pi$.
The transition function from
$\uhp$ to $\disc$ is given by
\begin{equation}
z(w)=i\,\frac{1-w}{1+w}\ .
\end{equation}
Primary fields of the CFT are denoted $\phi$ in the bulk, 
with conformal dimension $\Delta=2h$, and $\psi$ on the
boundary, with conformal dimension $h$. The identity
field is denoted by $\id$, both in the bulk and on the
boundary.
The basic correlators on the upper half-plane are
\begin{equation}
\begin{array}{rcl}
\coruhp{\psi(x)}&\;=\;&\delta_{\psi,\id}\coruhp{\id}\ , \\[10pt]
\coruhp{\psi(x_1)\psi(x_2)}&\;=\;&\tri{D}{\psi}{\psi}{\id}
\coruhp{\id}(x_1-x_2)^{-2h} \qquad \qquad \qquad \qquad (x_1>x_2)\ ,\\[10pt]
\coruhp{\phi(x+iy,x-iy)}&\;=\;&B_{\phi}{}^{\id}
\coruhp{\id}(2y)^{-\Delta} \qquad \qquad \qquad \qquad \qquad \qquad (y>0)\ ,\\[8pt]
\coruhp{\phi(x_1+iy_1,x_1-iy_1)\psi(x_2)}&\;=\;&
B_{\phi}{}^{\psi}\tri{D}{\psi}{\psi}{\id}\coruhp{\id}
(2y_1)^{-\Delta+h}\bigg((x_1-x_2)^2+y_1^2\bigg)^{-h}  \\
\vspace*{-0.7cm} 
& & \qquad \qquad \qquad \qquad \qquad  \qquad \qquad \qquad  \qquad \qquad (y_{1}>0)\ ,
\vspace*{0.3cm} \\[12pt]
\coruhp{\psi_1(x_1)\psi_2(x_2)\psi_3(x_3)}&\;=\;&
\tri{D}{\psi_2}{\psi_3}{\psi_1}\tri{D}{\psi_1}{\psi_1}{\id}
\coruhp{\id}(x_1-x_2)^{h_3-h_2-h_1}  \\[8pt]
&&\times (x_2-x_3)^{h_1-h_2-h_3}
(x_1-x_3)^{h_2-h_1-h_3} \nonumber \\
\vspace*{-0.7cm} 
& & \qquad \qquad \qquad \qquad \qquad  \qquad \qquad \quad \, \, \, \,  
\quad  (x_1>x_2>x_3)\ . \nonumber
\end{array}
\end{equation}
On the disc, these are
\begin{equation}
\begin{array}{rclr}
\cordisc{\psi(re^{i\theta})}&\;=\;&\delta_{\psi,\id}\cordisc{\id}\,,&\\[10pt]
\cordisc{\psi(e^{i\theta_1})\psi(e^{i\theta_2})}&\;=\;&\tri{D}{\psi}{\psi}{\id}
\cordisc{\id}\left|2\sin\frac{\theta_1-\theta_2}{2}\right|^{-2h}\ ,
&\\[10pt]
\cordisc{\phi(re^{i\theta},re^{-i\theta})}&\;=\;&B_{\phi}{}^{\id}
\cordisc{\id}(1-r^2)^{-\Delta}\ ,&\\[8pt]
\cordisc{\phi(re^{i\theta_1},re^{-i\theta_1})\psi(e^{i\theta_2})}&\;=\;&
B_{\phi}{}^{\psi}\tri{D}{\psi}{\psi}{\id}\cordisc{\id}
(1-r^2)^{-\Delta} \,
\big(\frac{1-r^2}{1-2r\cos(\theta_1-\theta_2)+r^2}\big)^{h}
\,,&\\[12pt]
\cordisc{\psi_1(e^{i\theta_1})\psi_2(e^{i\theta_2})\psi_3
(e^{i\theta_3})}&\;=\;&
\tri{D}{\psi_2}{\psi_3}{\psi_1}\tri{D}{\psi_1}{\psi_1}{\id}
\cordisc{\id}\left|2\sin\frac{\theta_1-\theta_2}{2}\right|^{h_3-h_2-h_1}&\\[8pt]
&&\times\left|2\sin\frac{\theta_2-\theta_3}{2}\right|^{h_1-h_2-h_3}
\left|2\sin\frac{\theta_1-\theta_3}{2}\right|^{h_2-h_1-h_3}\,.&
\end{array}
\end{equation}
In sections 2.1 and 3, we also need an expression for the correlator
\begin{equation}
\cor{\phi(z,\bar{z})\psi(x)\psi_(y)}_{\bf a}
\end{equation}
in the limit $m\rightarrow\infty$. This is a chiral
four-point function which can be computed from the
differential equation associated with the singular vector 
of the $(1,3)$ module at level 3,
\begin{equation}
\mathcal{N}=\left(L_{-3}-2h^{-1}L_{-2}L_{-1}+\frac{(m+1)^2}{2m(m-1)}
(L_{-1})^3\right)\phi\,.
\end{equation}
The general solution to the corresponding third order differential
equation in the cross-ratio $\eta$ can be obtained in the limit
$m\to\infty$. One can fix the solution by the asymptotic behaviour
when the bulk-field approaches the boundary, but as the dimensions of
the fields in the asymptotic channels differ by integers in the limit
$m\to\infty$, this is not the easiest method. A better approach
is to solve the differential equation for finite $m$ in an expansion
around $\eta =0$ where the channels can be clearly separated, and then
match the two sets of solutions in the limit $m\rightarrow\infty$.  Of
the three conformal blocks associated to the channels $(1,1)$,
$(1,3)$, and $(1,5)$, only the one involving the identity has a
nontrivial contribution for $m\rightarrow\infty$.  On the upper
half-plane, we find
\begin{equation}\label{UHPcorrelator}
\coruhp{\phi(z,\bar{z})\psi(x)\psi(y)}  = 
\frac{\eta_{1}^{4} -3 (\eta_{2}^{4}+\eta_{3}^{4})}{6 
(\eta_{1}\eta_{2}\eta_{3})^{2}}
B_{(1,3)}^{\;(1,1)}\tri{D}{(1,1)}{(1,3)}{(1,3)}
\tri{D}{(1,3)}{(1,3)}{(1,1)}
\coruhp{\id}\,(1+\mathcal{O}(m^{-1})) \ ,
\end{equation}
where 
\begin{align}
\eta_{1}&= (z-\bar{z}) (x-y)\ ,& \eta_{2} &= (z-x) (\bar{z}-y) \ ,&
\eta_{3}&= \eta_{2}-\eta_{1} = (z-y) (\bar{z}-x) \ .
\end{align}
Multiplying with $y^2$, this yields (\ref{correlator}) in the limit
$x\to 0$, $y\to \infty$. To perform the integral that leads to
(\ref{phipsipsi-contribution}), we need an expression for the correlator
on the disc,
\begin{align}
\cordisc{\phi(w,\bar{w})\psi(e^{i\theta_1})\psi(e^{i\theta_2})}=
\frac{\eta_{1}'^{4} -3 (\eta_{2}'^{4}+\eta_{3}'^{4})}{6
(\eta'_{1}\eta'_{2}\eta'_{3})^{2}}  
B_{(1,3)}^{\;(1,1)}\tri{D}{(1,1)}{(1,3)}{(1,3)}
\tri{D}{(1,3)}{(1,3)}{(1,1)}\coruhp{\id}\,(1+\mathcal{O}(m^{-1}))\ ,
\end{align}
where 
\begin{align}
\eta'_{1}&= (1-|w|^{2}) (e^{i\frac{\theta_{2}-\theta_{1}}{2}}
-e^{-i\frac{\theta_{2}-\theta_{1}}{2}}) \ ,&   
\eta'_{2} &= e^{i\frac{\theta_{1}-\theta_{2}}{2}} (w e^{-i\theta_{1}}-1)
(1-\bar{w}e^{i\theta_{2}})\ ,\\
\eta'_{3}&= \eta'_{2}-\eta'_{1} = e^{i\frac{\theta_{2}-\theta_{1}}{2}} 
(w e^{-i\theta_{2}}-1) (1-\bar{w}e^{i\theta_{1}}) \ .
\end{align}
Integrating this expression with our cut-offs $\epsilon$ on the boundary
and $\xi$ between bulk and boundary, we find
\begin{eqnarray}
&&\frac{2\pi}{3}B_{(1,3)}^{\;(1,1)}\tri{D}{(1,1)}{(1,3)}{(1,3)}\tri{D}{(1,3)}{(1,3)}{(1,1)}
\cordisc{\id}\,(1+\mathcal{O}(m^{-1}))\nonumber\\
&&\qquad\times\left(\left(\frac{3\pi (2-3\xi)}{2\xi}
+\mathcal{O}(\xi)\right)\frac{1}{\epsilon}
+\left(\frac{2\pi^2(3\xi-2)}{\xi}+\mathcal{O}(\xi)\right)\right.\\
&&\qquad\qquad\left.+\left(\frac{\pi}{\xi^2}-\frac{5\pi}{4\xi}+
\frac{\pi}{8}+\mathcal{O}(\xi)\right)\epsilon
+\mathcal{O}(\epsilon^2)\right)\,.\nonumber
\end{eqnarray}
The term proportional to $\epsilon^{-1}$ is subtracted when we compute
the connected correlation function. The $\epsilon$-independent part, on
the other hand, gives our result~(\ref{phipsipsi-contribution}).

\subsection{Semi-infinite cylinder}

Instead of considering the theory on the disc we now put it on a
semi-infinite cylinder by the conformal map
\begin{equation}
w=re^{i\theta} \mapsto v=-\log w = -\log r -i\theta =:v_{1}+iv_{2} \ .
\end{equation}
The coordinate $v_{1}$ runs from $0$ to $\infty$, whereas $v_{2}$ is
periodic with period $2\pi$. A bulk one-point function in these
coordinates reads
\begin{equation}
\langle \Phi (v,\bar{v})\rangle_{\text{C}} = B_{\phi}{}^{\id}\langle
\id \rangle \big(2\sinh v_{1} \big)^{-\Delta}\ .
\end{equation}
Let us consider the integral of a one-point function of a marginal
bulk field over the semi-infinite cylinder. To regularise the integral
we have to introduce a cutoff $\eta$ such that $v_{1}$ cannot come too 
close to the boundary. This leads to 
\begin{align}
I_{C} & = \int_{\eta}^{\infty} dv_{1}  \big(2\sinh v_{1} \big)^{-2} = \frac{1}{4} \big( -1+ \coth \eta\big)
= \frac{1}{4} \bigg(\frac{1}{\eta}-1+ \mathcal{O}(\eta )  \bigg) \ .
\end{align}
Comparing this with the disc integral $I_{3}$ in~\eqref{I3} that appeared
in the computation of the one-point contribution to the $g$-factor, we
see that we get coinciding results precisely when we relate the cutoffs
$\xi$ and $\eta$ as in~\eqref{etapres}. We conclude that on the
semi-infinite cylinder the natural cutoff $\eta$ is already the right
one. One could have expected this because here bulk and boundary
contributions should be disentangled as the boundary is not curved
with respect to the bulk metric, in contrast to the disc.\footnote{This
observation and the preceding computation resulted from discussions
with Anatoly Konechny.}

\section{Solutions of the RG equations}
\setcounter{equation}{0}

The set of RG equations in~\eqref{coupledRGeqs},
\begin{eqnarray}
\dot{\lambda} & = & \frac{4}{m} \lambda + 4 \pi \lambda^2  \ ,\label{lambdadot}\\
\dot{\mu}& = & \frac{2}{m} \mu + \frac{2\pi \alpha}{m} \lambda - 4 \pi \lambda \mu - 
\frac{4}{\alpha} \mu^2 
\label{mudot}
\end{eqnarray}
can be solved explicitly. First note that~\eqref{lambdadot} has the
general solution 
\begin{equation}
\lambda (t) =  \lambda_{*}\frac{C e^{\frac{4}{m}t}}{1+C e^{\frac{4}{m}t}}\ , \qquad
\lambda_{*} = - \frac{1}{\pi m} \ ,
\end{equation}
where $C $ is some constant. For the flows to connect $\lambda=0$ 
at $t\to -\infty$ and $\lambda =\lambda_{*}$ at $t\to +\infty$, we
have to take $C>0$. In that case $\lambda$ flows strictly 
monotonically, and we can parameterise 
\begin{equation}\label{muandg}
\mu(t)=\pi \alpha f(\lambda (t)) \ ,
\end{equation}
with some function $f(\lambda )$. Differentiating~\eqref{muandg} with respect to
$t$ and plugging in~\eqref{lambdadot} and~\eqref{mudot}, we obtain an
ordinary first order differential equation for $f(\lambda )$,
\begin{equation}
2\lambda (1+\pi m\lambda) f' (\lambda) = f (\lambda) + \lambda  -2\pi
m\lambda f (\lambda) -2\pi m f (\lambda)^{2} \ .
\end{equation}
This equation has the solution
\begin{equation}
f (\lambda) =
\lambda_{*}\frac{1+\chi
\sqrt{\frac{\lambda_{*}}{\lambda}-1}}{\frac{\lambda_{*}}{\lambda}-2
\bigg( 1+\chi \sqrt{\frac{\lambda_{*}}{\lambda}-1} \bigg) } \ ,
\end{equation}
with an arbitrary constant $\chi$. Let us discuss the different
solutions corresponding to different values of $\chi$ and their
properties in turn.

\begin{itemize}
\item For $\chi \to \infty$, the function $f(\lambda)$ becomes constant,
\begin{equation}
f(\lambda ) \equiv  -\frac{\lambda_{*}}{2} = \frac{1}{2\pi m} 
 \ \Longrightarrow \ \mu (t)\equiv 
\frac{\alpha}{2m} \ .
\end{equation}
This solution corresponds to the pure bulk flow from the fixed point
(I) to (II).
\item For $\chi$ finite, let us expand the function $f(\lambda
)$ for small (negative) $\lambda$,
\begin{equation}
f (\lambda) = \lambda_{*} \Big(  \chi \sqrt{\lambda/\lambda_{*}} + (1
+2 \chi^{2}) (\lambda/\lambda_{*})+\dotsb \Big) \ .
\end{equation} 
We see that for all finite $\chi$ we have $f(0)=0$, so that the
flow at $t=-\infty $ starts at $\lambda =\mu =0$. 
\item For $\chi=0$, $\mu$ grows at the same rate as $\lambda$ for
$t\to -\infty$. This corresponds to the situation when $\mu$ itself is
not turned on initially, but is just sourced by $\lambda$.
\item For finite $\chi\not= 0$, $\mu$ grows as
$\pi \alpha \lambda_{*}\chi\sqrt{C}e^{\frac{2}{m}t}$ for $t \to
-\infty$ which is the solution of the first order approximation
$\dot{\mu}=\frac{2}{m}\mu + \dotsb $ to the boundary RG
equation. $\chi$ thus determines how much and with which sign we
turn on the boundary field in the beginning, indeed we have
\begin{equation}\label{meaningofchi}
\chi = -\frac{1}{\pi \alpha \sqrt{|\lambda_{*}|}} \lim_{t \to -\infty}
\frac{\mu (t)}{\sqrt{|\lambda (t) |}} \ . 
\end{equation}
\item For any finite $\chi$ the function $f(\lambda )$ has a pole at
\begin{equation}
\lambda_{0} = \frac{\lambda_{*}}{2} 
\bigg(1 - \frac{\chi}{\sqrt{1+\chi^{2}}} \bigg) \ ,
\end{equation}
at which the function $f(\lambda )$ diverges to $-\infty$ when
$\lambda /\lambda_{*}$ approaches $\lambda_{0}/\lambda_{*}$ from
below. Note that the perturbative fixed points (I) and (II) at $\mu
=\alpha/2m$ correspond to $f=-\lambda_{*}/2 > 0$, so the solution always
runs away from the perturbative fixed point for large $t$.   
\item For finite $\chi<0$, the function $f$ first develops 
towards positive values, then reaches its most positive value 
\begin{equation}
f_{\text{max}} = f (\lambda_{\text{max}}) =
-\frac{\lambda_{*}}{2}\bigg(1-\frac{1}{\sqrt{1+\chi^{2}}} \bigg) \quad
\text{at} \quad \lambda_{\text{max}} =
\frac{\lambda_{*}}{2}\bigg(1-\frac{1}{\sqrt{1+\chi^{2}}} \bigg) \ ,
\end{equation}
and then runs towards $-\infty$. The maximal value $f_{\text{max}}$
approaches $-\lambda_{*}/2$ for $\chi\to -\infty$, so we can come
arbitrarily close to the perturbative fixed-point at $\mu=\alpha/2m$ by
tuning $\chi$ to large negative values.
\end{itemize}
\smallskip

\noindent
Similarly, the solution to the RG equation (\ref{RGeqmiddle}) is of the form
\begin{equation}\label{RGsolutionmiddle}
\mu=\pi a_2 f(\lambda(t))\ , \qquad
f(\lambda)=-\frac{m\lambda}{2}+\sqrt{-\lambda}\sqrt{1+m\pi\lambda}\,\chi \ ,
\end{equation}
where $\chi$ is again an arbitrary real parameter. 

\end{appendix}

\end{document}